\documentclass[preprint,12pt]{elsarticle}
\usepackage{amssymb}
\usepackage{a4wide}
\usepackage[utf8]{inputenc}
\usepackage[T1]{fontenc}   
\usepackage{hyperref}      
\usepackage{url}           
\usepackage{booktabs}      
\usepackage{amsfonts}
\usepackage{amsmath}      
\usepackage{nicefrac}      
\usepackage{microtype}     
\usepackage{cleveref}      
\usepackage{graphicx}
\graphicspath{{./figs/}}
\usepackage{doi}
\usepackage{subcaption}
\usepackage{bm}
\usepackage{algorithm}
\usepackage{algpseudocode}
\usepackage{setspace}
\usepackage[inkscapepath=svgsubdir]{svg}

\newcommand{\B}[1]{\bm{#1}}

\newcommand{\tran}{^{\mkern-1.5mu\mathsf{T}}}

\newcommand{\fw}{.9}

\usepackage{tikz}
\usetikzlibrary{trees}
\usetikzlibrary{shadows.blur}
\usetikzlibrary{shapes,arrows.meta,positioning}

\journal{Journal of Sound Vibration}

\begin{document}

\begin{frontmatter}

\title{Full-scale modal testing of a Hawk T1A aircraft for benchmarking vibration-based methods}

\author[uos_add,eth_add]{Marcus Haywood-Alexander\corref{cor1}\fnref{fn1}}
\ead{m.haywood@ethz.ch}
\author[uos_add]{Robin S.\ Mills}
\ead{robin.mills@sheffield.ac.uk}
\author[uos_add]{Max D.\ Champneys}
\ead{max.champneys@sheffield.ac.uk}
\author[uos_add]{Matthew R.\ Jones}
\ead{matthew.r.jones@sheffield.ac.uk}
\author[uos_add]{Matthew S.\ Bonney}
\ead{m.bonney@sheffield.ac.uk}
\author[uos_add,ati_add]{David Wagg}
\ead{david.wagg@sheffield.ac.uk}
\author[uos_add]{Timothy J. Rogers}
\ead{tim.rogers@sheffield.ac.uk}

\cortext[cor1]{Corresponding author}
\fntext[fn1]{The work published here was undertaken at the University of Sheffield, but this author is now affiliated at ETH Zurich.}

\affiliation[uos_add]{organization={Dynamics Research Group, University of Sheffield},
            addressline={Mappin Street}, 
            city={Sheffield},
            postcode={S1 4ET},
            country={UK}}

\affiliation[eth_add]{organization={Chair of Structural Mechanics and Monitoring, ETH Zurich},
            addressline={Wolfgang-Pauli-Strasse}, 
            city={Zurich},
            postcode={8049},
            country={Switzerland}}

\affiliation[ati_add]{organization={The Alan Turing Institute},
            city={London},
            postcode={NW1 2DB},
            country={UK}}

\begin{abstract}
Research developments for structural dynamics in the fields of design, system identification and structural health monitoring (SHM) have dramatically expanded the bounds of what can be learned from measured vibration data. However, significant challenges remain in the tasks of identification, prediction and evaluation of full-scale structures. A significant aid in the roadmap to the application of cutting-edge methods to the demands of in-service engineering structures, is the development of comprehensive benchmark datasets. With the aim of developing a useful and worthwhile benchmark dataset for structural dynamics, an extensive testing campaign is presented here. This recent campaign was performed on a decommissioned BAE system Hawk T1A aircraft at the Laboratory for Verification and Validation (LVV) in Sheffield. The aim of this paper is to present the dataset, providing details on the structure, experimental design, and data acquired. The collected data is made freely and openly available with the intention that it serve as a benchmark dataset for challenges in full-scale structural dynamics. Here, the details pertaining to two test phases (frequency and time domain) are presented. So as to ensure that the presented dataset is able to function as a benchmark, some baseline-level results are additionally presented for the tasks of identification and prediction, using standard approaches. It is envisaged that advanced methodologies will demonstrate superiority by favourable comparison with the results presented here. Finally, some dataset-specific challenges are described, with a view to form a hierarchy of tasks and frame discussion over their relative difficulty.  
\end{abstract}

\begin{keyword} Modal testing \sep Hawk T1A \sep Full-scale dynamic testing \sep Shaker test \sep SIMO testing \sep Aircraft dynamics \end{keyword}

\end{frontmatter}


\section{Introduction}
\label{sec:introduction}

How a structure behaves under dynamic loading is an imperative part of engineering design, operation, and maintenance. %
In both experimental and operational contexts, the measuring and monitoring of modal characteristics can provide vital information on the structure \cite{craig2006fundamentals,hodges2011introduction}. %
These modal characteristics can be used to, either directly or indirectly, infer the physical properties of the system. %
One such use of this information is to determine the existence of damage in important locations on the structure. %
When damage, such as cracks or disbonds, occur on the structure, this corresponds to a localised loss of stiffness, which manifests as a change in the eigenfrequencies of the system. %
However, structural nonlinearities can also cause apparent changes to modal properties therefore causing false-positive or false-negative detection \cite{hemez2001review}.  %

Using a well-defined model, it is possible to generate estimates of physical parameters using analytical, or purely-physics modelling (derived from first-principles analysis of elastic deformation theory). The presence of nonlinearity significantly complicates this task. The modelling of \emph{known} nonlinearities is in itself a difficult task \cite{kerschen2006past,worden2019nonlinearity}, however, when the nonlinearities are unknown, or ill-prescribed, in the model, this presents an even more difficult challenge. %
This, along with the challenge of learning model parameters, has led to the increased development of data-driven techniques to estimate or learn the models and parameters \cite{rogers2017grey,worden2018evolutionary,fuentes2021equation}. %
As these algorithms are developed, it is important that they are consistently assessed on meaningful examples, or benchmarks. %
The meaningfulness, as we define it here, of such benchmarks can be described by two characteristics; they present a reasonable challenge in that they that cannot be trivially solved by simple, well-established approaches, and that they provide data from a variety of loading or operating conditions.

In this work, we present a dataset with the aim to fulfil the requirements of a meaningful benchmark for structural modal analysis. %
In order to do so, experimental acceleration data was collected from a BAE system Hawk T1A aircraft undergoing a variety of dynamic input excitation, with simulated damage scenarios. %
The aim of the dataset is to provide a challenging benchmark for structural dynamics, in fields such as system identification and structural health monitoring. %
As well as presenting the data, an ancillary contribution of this paper is to also present a baseline standard in the tasks of system identification and structural health monitoring. %
The techniques presented here do not necessarily represent the state of the art in these two tasks but instead offer a first analysis; indicative of approaches that might be employed before more advanced techniques are sought. %
The rationale behind these choices is to establish how much insight can be gained from basic techniques and to act as a baseline against which future approaches can be assessed. %
A further motivation is to qualitatively establish the difficulty of selected system identification and structural health monitoring tasks on the Hawk data. %
This is an important undertaking as it highlights the issues presented when advanced machine learning (ML) methods are applied to case studies that can be solved by more simplistic, robust, and interpretable methods. %

The Hawk dataset is freely provided with the motivation that it might be used to develop new algorithms for structural dynamics. %
The data is available in open \emph{.hd5} format, and access is provided via a simple Python API, which is also provided.\footnote{All data and documentation for the API is available at \url{https://doi.org/10.15131/shef.data.22710040.v1}.} %

The remainder of this paper is arranged as follows. %
In \Cref{sec:structure}, details of the aircraft structure are given, showing the boundary conditions and general information. %
This is followed by \Cref{sec:experiment}, where details on the experiment and data collection are provided, including information on the full experimental regime, data collected, control system, hardware, and acquisition. %
At this point the paper moves on to the baseline tasks applied to the dataset. %
\Cref{sec:system_id} presents results of a series of system identification tasks applied to the dataset, both in the frequency and time domain. %
Following this, results from a structural health monitoring perspective are given in \Cref{sec:shm}. %
Finally, in \Cref{sec:challenges} some observations are provided on the challenges of the dataset, in order to showcase the use and meaningfulness of the data. %

\section{Structure}
\label{sec:structure}

\begin{figure}
	\centering
	\begin{subfigure}[b]{0.49\textwidth}
		\centering
		\includegraphics[width=\textwidth]{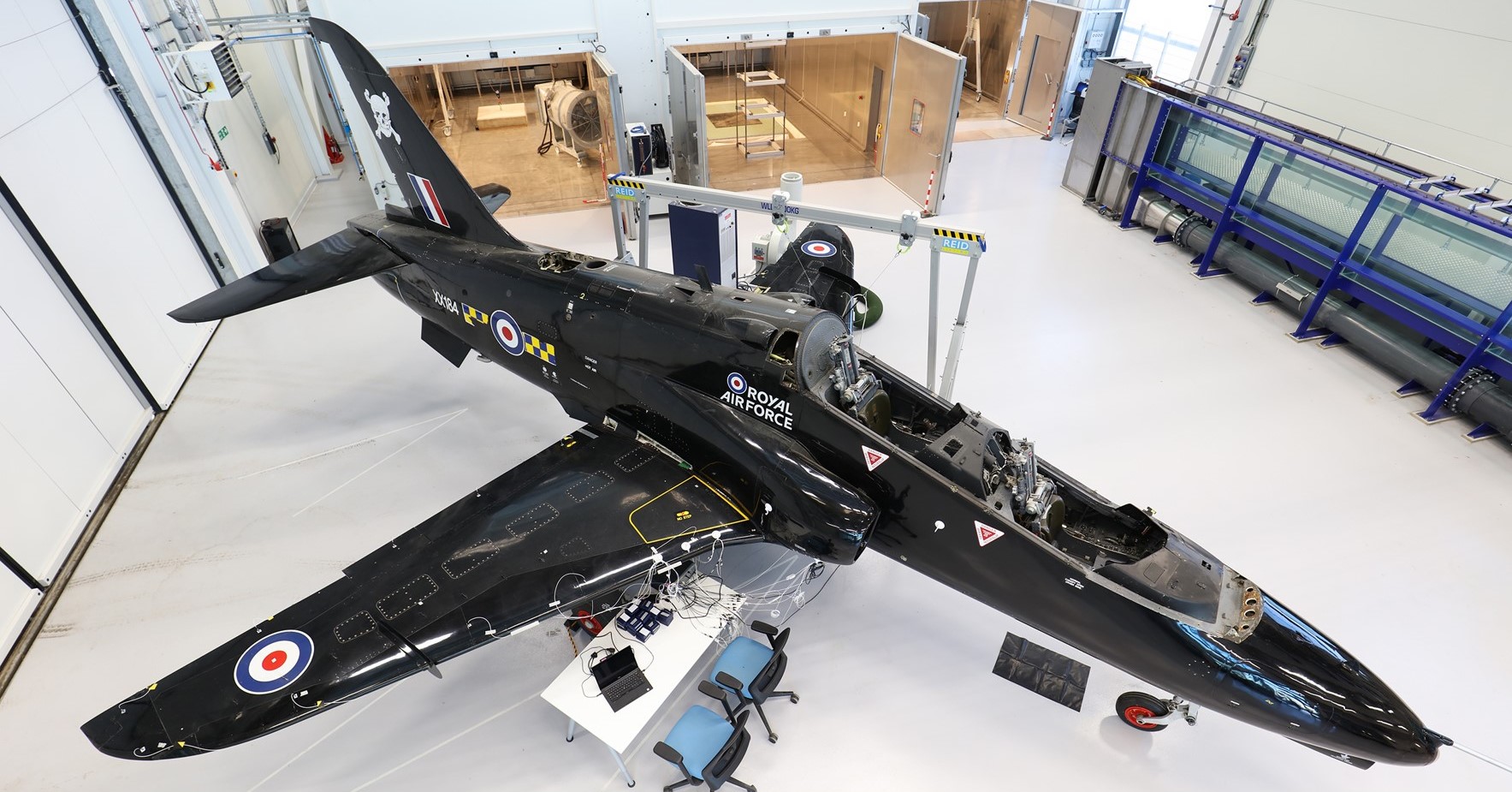}
		\caption{Image of aircraft.}
		\label{fig:aircraft_diagram}
	\end{subfigure}
	\hfill
	\begin{subfigure}[b]{0.49\textwidth}
		\centering
		\includegraphics[width=\textwidth]{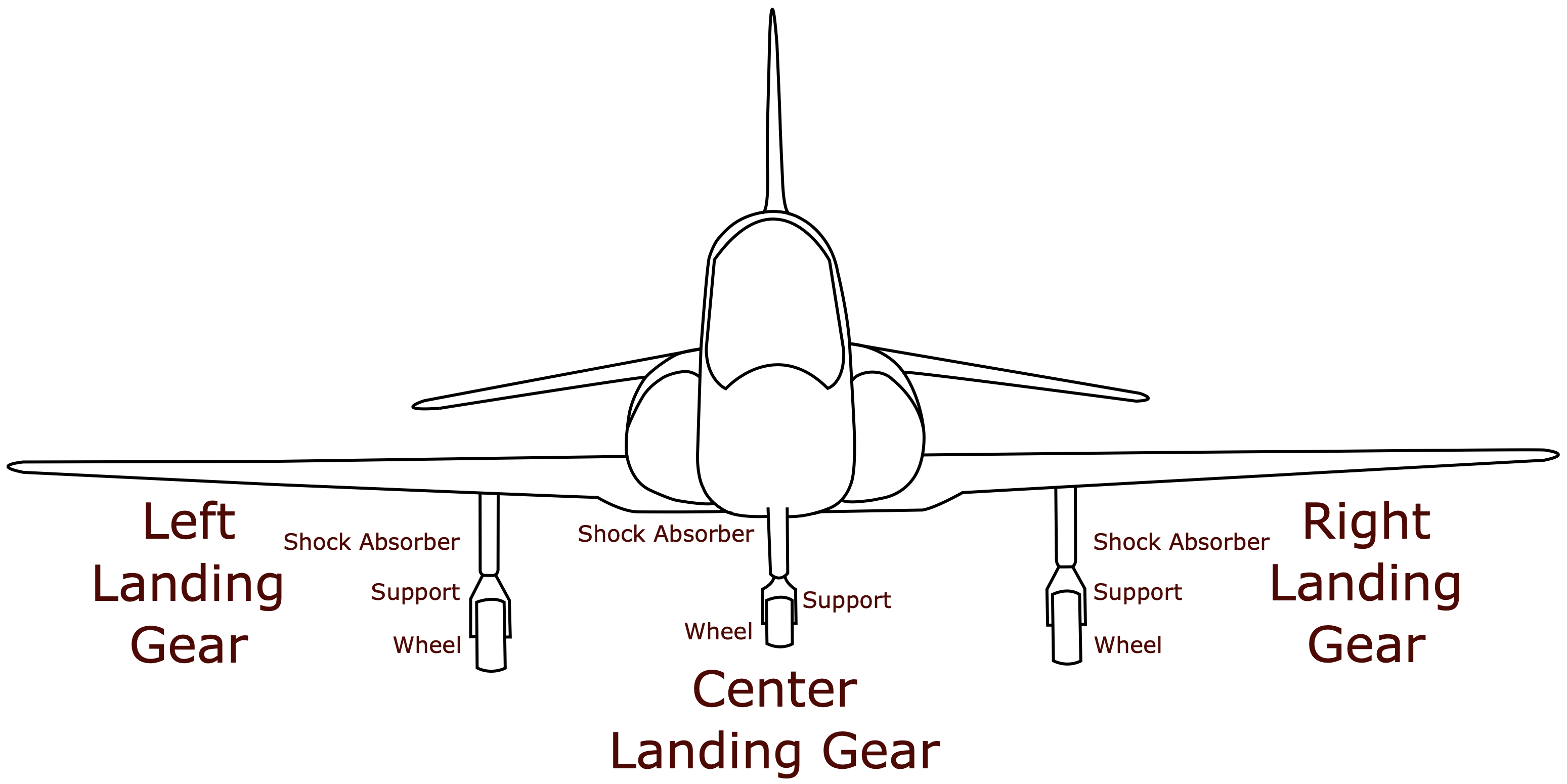}
		\caption{Diagram showing front view of aircraft.}
		\label{fig:aircraft_real}
	\end{subfigure}
    \caption{Images of BAE systems Hawk T1A aircraft.}
    \label{fig:hawk-visual}
\end{figure}

The structure used in this dataset is a BAE systems Hawk T1A aircraft \cite{fraser2011hawk}.
This is a real, decommissioned aircraft, that was used for advanced training in the RAF. %
The aircraft was tested at the Laboratory for Verification and Validation (LVV) in Sheffield, England. %
For this dataset, only the starboard wing was measured, as well as supplementary measurements at the top of the aircraft and the exhaust, more details on the exact locations of the sensors will be given later in \Cref{sec:setup_hardware}, \Cref{fig:sensor_locs}. %
The aircraft was positioned resting on its wheels on the ground. %
Therefore, between the main body and the ground are the wheels, supports and hydraulic shock absorbers. %
All these substructures are joined with a variety of fixing methods and types, which are not detailed here. %
\Cref{fig:hawk-visual} shows both an image of the real aircraft,  as well as a schematic, indicating the boundary conditions of the wing. %

\section{Experiment}
\label{sec:experiment}

The aim of this work was to generate modal analysis data for the starboard wing of the Hawk aircraft. %
To do so, the wing was excited with a shaker, and the acceleration response over the upper and lower surfaces of the wing were measured using accelerometers, and processed to return the response in the frequency domain. %
This section will detail all the experiment details, beginning with the testing regime, the experiment setup and hardware, and then acquisition and processing details. %

\subsection{Regime outline}

\begin{figure}
    \centering
    \includegraphics[width=1.0\textwidth]{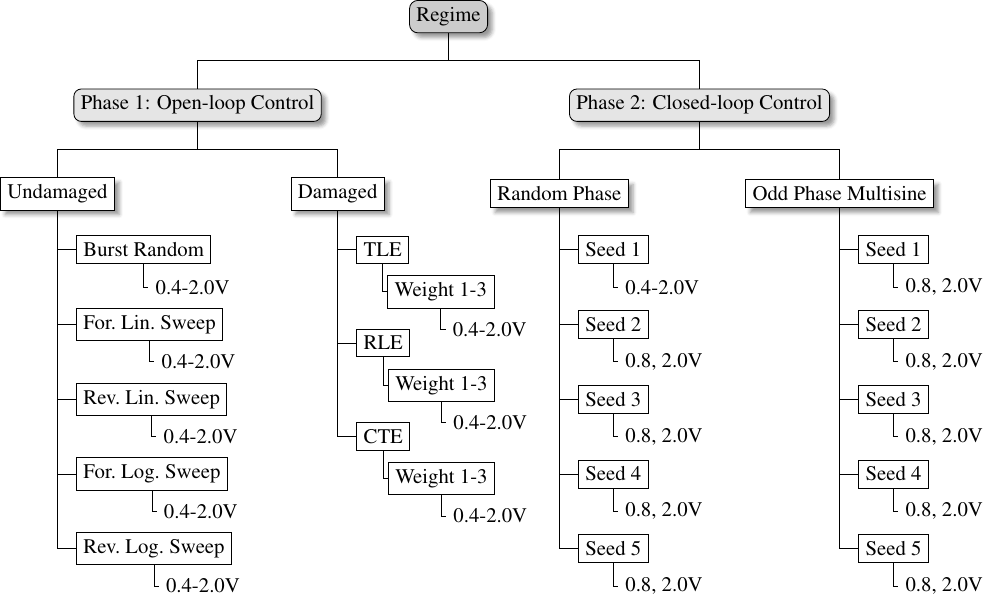}
    \caption{Testing regime for dataset. For the damaged tests, the acronyms represent the locations of the added mass, where TLE, RLE, and CTE stand for \emph{tip-leading-edge}, \emph{root-leading-edge}, and \emph{central-trailing-edge}, respectively. The weight numbers are ID-representations, the value of mass is provided in the experiment details.}
    \label{fig:regime_hier}
\end{figure}

The total experimental regime was split into two phases; open-loop control using off-the-shelf hardware, and a custom-built, closed-loop control system. %
Specific details on the control system and hardware will be given in \Cref{sec:setup_hardware}. %
Each of these experimental regimes followed a hierarchical structure based on the aims of the experiment; this is shown in \Cref{fig:regime_hier}. %
The lowest level of this hierarchy is herein referred to as a `test', and each test was performed with 10 repetitions. %
For clarification; a repetition is where the run is finished, allowed to reset, and then repeated. %
This repeat style was done to allow for quantification and mitigation of experimental error. %

\subsubsection{Phase 1: Open-loop control}
In the first phase, there were two main sub-regimes; the nominally-undamaged state, and the pseudo-damaged state. %
In the first case, the aircraft was left in its assumed healthy (free from significant structural damage) condition, and in the second case, damage was `simulated' by adding masses of different weights in specified locations. %
For the undamaged tests, the method of forcing input was set to either burst random, or forward/reverse sine sweeps in the linear/logarithmic scale. %
The burst random uses a white noise forcing input, the amplitude of which is ramped at the beginning and end of the burst. %
For the sine sweeps, the forcing input is a sine wave signal of which the frequency increases during the time period of the measurement. %
Forward/reverse refers to the direction of the frequency sweep, and linear/logarithmic refers to the pattern of the sweep. %
This means that for forward logarithmic sweeps, more time is spent exciting the lower frequencies, and for reverse logarithmic sweeps, more time is spent exciting the higher frequencies. %
For each of the excitation types, the amplitude of the forcing was increased by increasing the output voltage in increments of 0.2V. %

For the damaged cases in Phase 1, only burst random excitation was used, and to simulate a range of damage cases, the weight and location of the added masses was changed. %
In \Cref{fig:regime_hier}, TLE, RLE, and CTE refer to these locations, and stand for \emph{tip-leading-edge}, \emph{root-leading-edge}, and \emph{central-trailing-edge}, respectively. %
The locations on the wing are shown in \Cref{fig:sensor_locs}, and exact details are given in the accompanying technical report \cite{HaywoodAlexander2023}. %
Much like the undamaged tests, the amplitude of the forcing was increased by increasing the voltage, but this time in increments of 0.4V. %

\subsubsection{Phase 2: Closed-loop control}

The second phase of testing is split into sub-regimes based on the target signal types, by iteratively generating the signal passed to the shaker. %
More details on the control algorithm are given in \Cref{sec:control_algo}, along with the definition of each of the actuation signal types. %
To determine information on the effect of randomness, each of these signal types are tested with different initial seeds. %
To increase the amplitude of the forcing, as this is a closed-loop control which uses the measured amplitude, changing the voltage output will simply cause the control algorithm to decrease the signal level. %
Therefore, instead the amplitude of the target spectra were increased to corresponding levels. %
It is important to note, these do not necessary correspond exactly to the levels used in Phase 1, they are instead values which return amplitudes of a similar spectral density. %

\subsection{Setup and hardware}
\label{sec:setup_hardware}

To measure the structural response, PCB Piezotronics accelerometers were used with a nominal sensitivity of 10 mV/g. %
When converting voltage readings to acceleration values, the sensor-specific sensitivities, accurate to 0.01 mV/g, were used. %
A total of 54 sensors were placed on the wing, which are placed in three lines on the upper surface, and two lines on the lower surface of the wing, where a line runs from the root to the tip. %
As well as these 5 strips, additional sensors were placed to provide information at boundary condition locations of the root and the open panel for the landing wheel. %
The force was applied to the lower surface of the wing using a Tira TV 51140-MOSP shaker, attached directly to a PCB Piezotronics 208C02 force transducer with a sensitivity of 11.11 mV/N. %

\begin{figure}[h!]
    \centering
    \includegraphics[width=0.8\textwidth]{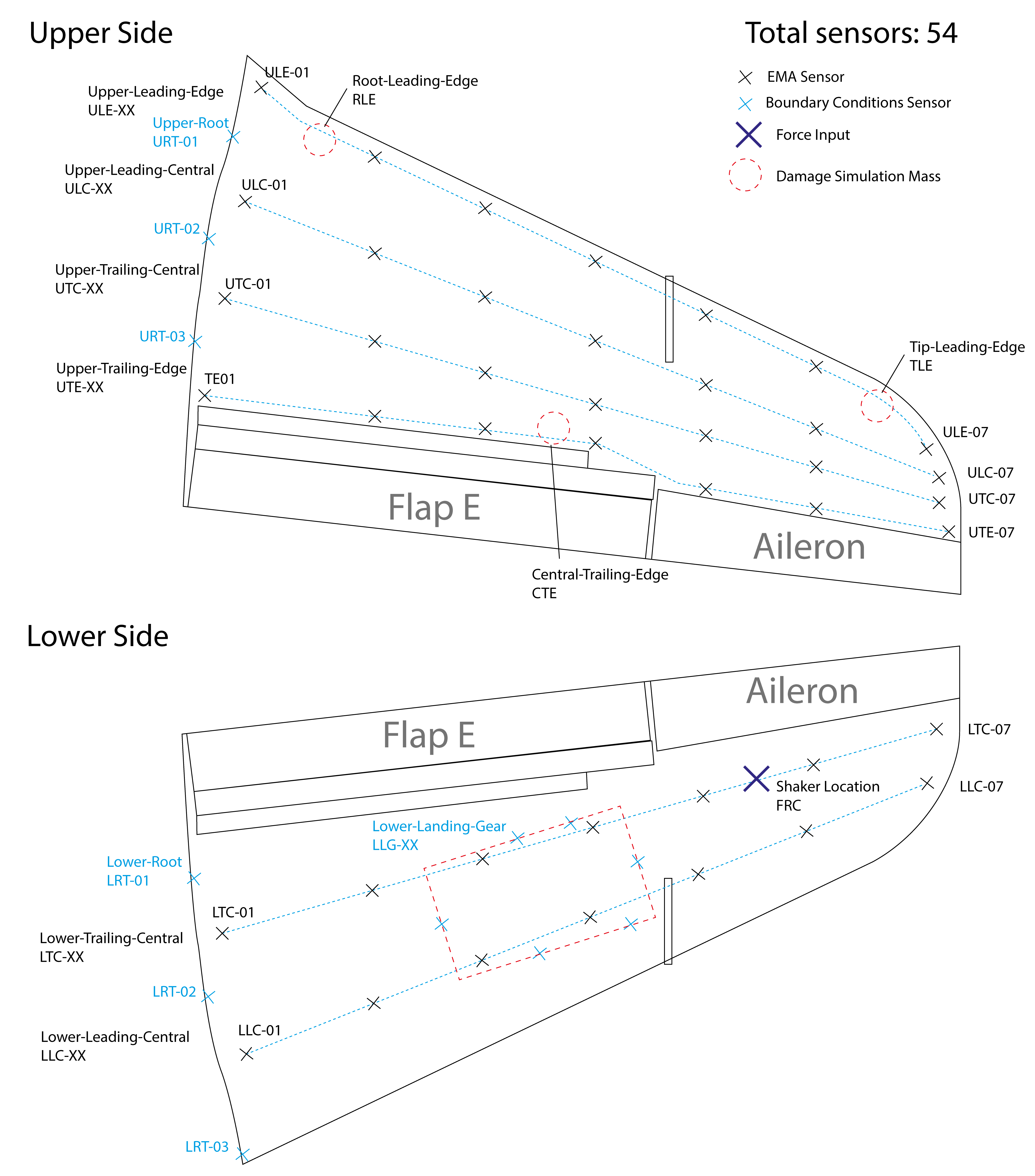}
    \caption{Sensor layout on the wing, also indicating the locations of the shaker and additional masses.}
    \label{fig:sensor_locs}
\end{figure}

A diagram showing the locations of the sensors, forcing, and masses is given in \Cref{fig:sensor_locs}. %
The locations of the sensors in this diagram are the nominal locations, but may have been slightly shifted in order to avoid components of the wing, exact coordinates are provided in the data and the technical report \cite{HaywoodAlexander2023}. %
The black sensor locations indicate those placed for the purposes of experimental modal analysis of the general structure. %
The blue sensor locations indicate those placed for the collection of data relating to the boundary conditions. %
Also shown are the shaker attachment location, indicating where the excitation force is applied, and the locations of the added masses and their corresponding ID tags. %

For the open-loop control, a Siemens PLM LMS SCADAS unit was used to generate the forcing signal, and simultaneously measure the acceleration responses. %
For the closed-loop control system, National Instruments hardware was used, along with LabView, to develop a frequency-domain control system, targeting the force signal at the location of the shaker. %
A table of the full hardware is shown in \Cref{tab:CL_hardware}. %

\begin{table}[h!]
    \centering
    \begin{tabular}{l | l | l}
        Name & Quantity & Use \\
        \hline
        NI-cDAQ-9179 & 2 & Chassis for total system\\
        NI-9232 & 20 & Measures voltage signals from accelerometers\\
        NI-9260 & 1 & Outputs voltage signal\\
    \end{tabular}
    \caption{Hardware components for closed-loop control system.}
    \label{tab:CL_hardware}
\end{table}

\subsubsection{Control algorithm}
\label{sec:control_algo}

\begin{figure}[h!]
    \centering
    \includegraphics[width=0.8\textwidth]{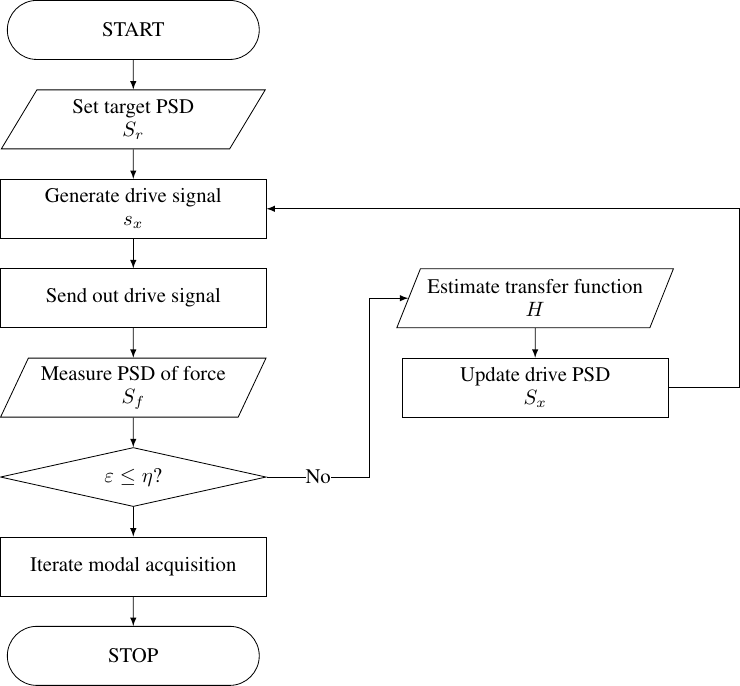}
    \caption{Control algorithm process flow, indicating the total workflow for one \emph{repeat} of a test.}
    \label{fig:control_flow}
\end{figure}

In the second phase of testing, a custom algorithm was implemented to control the actuation signal in the frequency domain. %
This allows the use of pre-prescribed forcing types, for both amplitude and phase of the frequencies. %
The overall control flow is shown in \Cref{fig:control_flow}, showing the process which covers one repeat of a test. %
The general aim of the algorithm is to control the output of the shaker to the structure in the frequency domain. %
For this work, the frequency domain information is defined and calculated as the power-spectral-density (PSD), and contains both magnitude and phase information. %
The target PSD of this force is denoted $S_r(\omega)$, and the measured PSD from the transducer is denoted as $S_f(\omega)$. %
The algorithm estimates a model of the mapping between the driving signal, in terms of voltage, to the applied force. %
This is achieved by assuming a linear transfer function,
\begin{equation}
    |F(\omega)| = |H(\omega)||S_d(\omega)|
\end{equation}
where $F(\omega)$ is the force applied to the structure, $H(\omega)$ is the transfer function, and $S_d(\omega)$ is the drive signal in volts, all in the frequency domain as functions of the circular frequency $\omega$. %

The target actuation $S_r(\omega)$ is defined in terms of the PSD of the force applied. Details of how these target signals are generated are provided in \Cref{sec:actuation_descrip}. %
At the first step of the control algorithm, the drive PSD $S_d(\omega)$ is simply set to the reference (i.e.\ $|H| = 1$). %
At each iteration, the drive signal $s_x(t)$, is calculated using an inverse Fourier transform,
\begin{equation}
    s_d(t) = \mathcal{F}^{-1}(|S_d(\omega)|)
    \label{eq:drive_sig_gen}
\end{equation}
When the actuation signals were generated, a linear amplitude ramp was applied to window the signal in the time domain, to reduce shock to the structure and equipment. %
This signal is sent out from the control system, and the PSD of the force, $S_{f}(\omega)$ is measured. %
This measurement is then used to update the estimate of the linear transfer function using,
\begin{equation}
    |H(\omega)| = \alpha\sqrt{\frac{S_r(\omega)}{S_f(\omega)}}
\end{equation}
where $0.0 \geq \alpha \leq 1.0$ is a proportional weighting factor which is included to slowly iterate the estimated transfer function to mitigate potential out-of-control issues, or overloading of the system. %
The value of this is chosen by the control system designer, and in this case was chosen to be 0.8. %
Then, the drive PSD is updated, and subsequently the drive signal using \Cref{eq:drive_sig_gen},
\begin{equation}
    |S_d(\omega)| = \frac{|S_r(\omega)|}{|H(\omega)|}
\end{equation}
During the drive updating, the target PSD error is defined as the mean squared error,
\begin{equation}
    \varepsilon = \frac{1}{N}\sum_{i=1}^N \left( |S_f(\omega^{(i)})| - |S_r(\omega^{(i)})| \right)^2
\end{equation}
where $N$ is the number of frequency points in the discretised frequency domain. %
Once the error is below a specified target, $\eta$, the latest drive PSD is used for the acquisition stage. %
For this work, the target error was set to 0.01. %

\subsubsection{Frequency Domain Actuation Types}
\label{sec:actuation_descrip}

This subsection aims to explain the difference between the actuation types in Phase 2, which are defined in the frequency domain. %
Both actuation types generate the time domain signal as a sum of the discrete frequencies represented by their definitions in terms of power-spectral-density. %
The actuation is defined using a vector of frequencies, $\pmb{\omega}$, with corresponding vectors of amplitude, $\mathbf{A}$, and phase, $\pmb{\phi}$. %
Where $N$ is the length of these vectors, and is the same as the PSD definitions in the above subsection, the discrete time signal, $s(t)$ is defined as as scalar function of the discrete time points,
\begin{equation}
  s(t) = \sum_{i=1}^{N} A_i\sin(\omega_i t + \phi_i)
\end{equation}

The vector of frequencies is chosen based on the bandwidth of modes of interest, and its discretisation is aligned with the acquisition parameters. %
Generally, the vector $\mathbf{A}$ is designed to `ramp up and ramp down' the amplitude with respect to the frequencies. %
This is done in order to maintain periodicity in the time domain signal to avoid shock in the physical system. %
The phase is generated by randomly sampling from a uniform distribution between 0 and $2\pi$. %
Using these generation methods with no modification forms the \textbf{random phase} actuation type. %
However, for the \textbf{odd phase} actuation, every \emph{other} element in the vectors $\mathbf{A}$ and $\pmb{\phi}$ are set to zero,
\begin{equation}
  A_i=0, ~~~~ \phi_i=0, ~~\text{for}~~ i= 2n-1~~\text{for}~~0\leq n \leq N/2
\end{equation}

\begin{figure}[h!]
    \centering
    \includegraphics[width=0.9\textwidth]{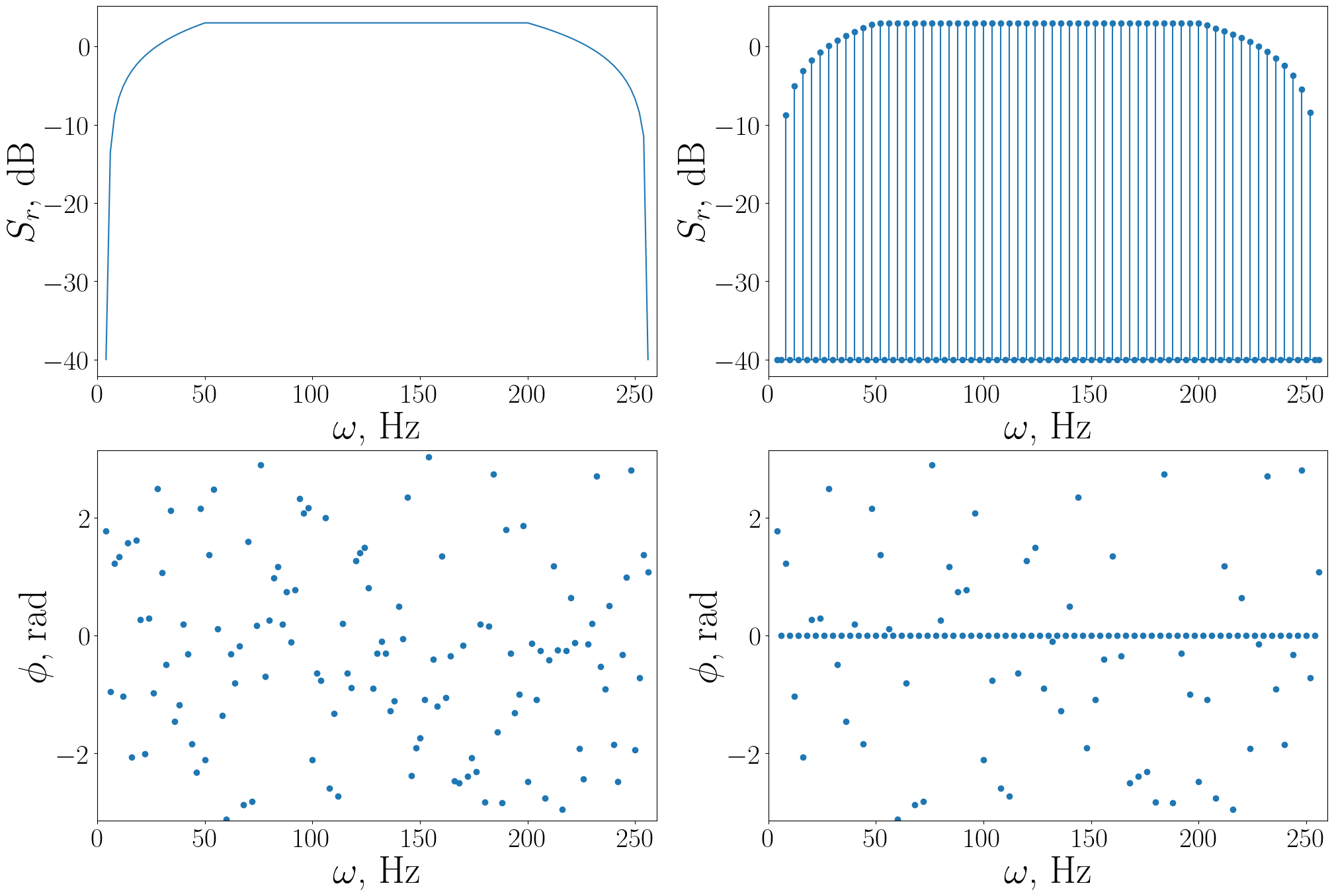}
    \caption{Examples of the target driver PSDs for (left) random phase and (right) odd phase multisine types. The top plots show the target PSD power in terms of dB, and the bottom plots show the corresponding phase.}
    \label{fig:sr_example}
\end{figure}

\subsection{Acquisition and Processing}

For all the tests, a number of acquisition parameters (listed in \Cref{tab:acqu_params}) were kept constant. %
The number of samples indicates the number of sampling points taken in the time domain. %
The burst/sweep time represents the portion of the measured signal in which the random burst or the sine sweep takes place; i.e.\ for 90\% burst time, the first and last 5\% of the acquisition time has no actuation. %

\begin{table}[h!]
    \centering
    \begin{tabular}{l | l || l | l}
        \multicolumn{2}{c}{Phase 1} & \multicolumn{2}{c}{Phase 2} \\
        \hline
        \hline
        Parameter & Value & Parameter & Value \\
        \hline
        Sample Rate & 512 Hz & Sample Rate & 2048 Hz \\
        Number of samples & 16,384 & Number of samples & 32,768 \\
        Burst/sweep time & 90\% & Burst time & 90\% \\
        Amplifier level & 50\% & Amplifier level & 50\% \\
        Repetitions & 10 & Repetitions & 10 \\
    \end{tabular}
    \caption{Acquisition parameters for each test phase.}
    \label{tab:acqu_params}
\end{table}

The data is then saved with identifiers which relate to the test series as shown in \Cref{fig:regime_hier}. %
The sub-regime identifiers are shown in \Cref{tab:exp_data_identifiers}, where the placeholder `X' in the identifiers are the test numbers identifiers. %
These are integer values which correspond to the voltage level, added mass weight, and/or seed used. %
Further details of the exact labels for each individual test can be found in the technical report \cite{HaywoodAlexander2023}. %

\begin{table}[h!]
    \centering
    \begin{tabular}{l|l||l|l}
        \multicolumn{2}{c||}{Phase 1} & \multicolumn{2}{c}{Phase 2} \\
        \hline
        \hline
        Undamaged & Damaged & Random Phase & Odd Phase Multisine \\
        \hline
        BR\_AR\_X & DS\_TLE\_X & RPH\_AR\_X & ORP\_VS\_X \\
        FSS\_AR\_X & DS\_RLE\_X & FLS\_AR\_X & \\
        RSS\_AR\_X & DS\_CTE\_X & RLS\_AR\_X & \\
        FLS\_AR\_X & & & \\
        RLS\_AR\_X & & & 
    \end{tabular}
    \caption{Test identifiers used for naming data in acquisition and processing.}
    \label{tab:exp_data_identifiers}
\end{table}

\subsection{Dataset structure}

In order to keep this paper self-contained, a brief description of the dataset structure is included here.  
All data is stored in the open-source \emph{hierarchical data format} (hd5). The hd5 format can be accessed using several open-source tools, but the authors also provide a python interface.\footnote{A python API for the hawk data is available at \url{https://github.com/MDCHAMP/hawk-data}} The hd5 format essentially mimics the tree-like file structure of modern computers. Data can be accessed by specifying the path to the desired sensor channel and data series. The general form of the path for the Hawk SBW test data is:

\begin{center}
\verb|/SBW/{test phase}/{test series}/{repetition}/{sensor}/{signal}|
\end{center}

where SBW (starboard wing test) is the overall test campaign the bracketed quantities refer to the following optional fields:

\begin{itemize}
    \item \verb|test phase| $\in$ \verb|{LMS, NI}|: Refers to the frequency domain and time domain data respectively, as per Table \ref{tab:acqu_params} and Figure \ref{fig:regime_hier}.
    \item \verb|test series| $\in$ \verb|{BR_AR, DS_TLE, ...}| (Phase 1) and $\in$ \verb|{RPH_AR, ORP_VS}| (Phase 2): Test series identifier (see Figure \ref{fig:regime_hier} for definitions).
    \item \verb|repetition| $\in$ \verb|{01, 02, ...}|: Test repetition number. For each test series there are nominally 10 repetitions, although for some tests fewer repetitions are available.
    \item \verb|sensor| $\in$ \verb|{FRC, LLE-01, ..., LLC-07}|: Sensor location (see Figure \ref{fig:sensor_locs} for definitions).
    \item \verb|signal| $\in$ \verb|{(coherenceSpectrum, spectra, frf}| (Phase 1) and $\in$ \verb|{acc, (force)}| (Phase 2): Desired signal. Note that frequency-response-functions (FRFs) are only available for output channels in the phase 1 tests.

\end{itemize}

The self-describing nature of the .hd5 format allows metadata to be stored alongside the collected series. Test data related to sensor position, sensitivities, units and test parameters (added mass, excitation level etc.) can all be accessed by querying the metadata fields at each level of the .hd5 hierarchy. Additional detail regarding the dataset structure and interface can be found at \url{https://doi.org/10.15131/shef.data.22710040.v1}.

\section{System identification}
\label{sec:system_id}
With the description of the experimental campaign complete, attention can now be turned to establishing a benchmark for the Hawk data in two common tasks in structural dynamics analysis \emph{system identification} and \emph{structural health monitoring}. It is envisaged that the results here will act as a baseline level against which more advanced methods might be compared in future work.

One of the key challenges in structural dynamics is obtaining accurate models of system dynamics. For structures that can be considered predominantly linear, the default framework for modelling dynamics is that of \emph{modal analysis} \cite{ewins2009modal}. Modal analysis is familiar to many engineers as a method of decomposing system dynamics into a number of eigenfrequencies and eigenfunctions that encode the response of the system to any input. These properties are invariant to any excitation and can be conveniently expressed in terms of a discrete number of interpretable properties; the natural frequencies $\omega_r$, damping ratios $\zeta_r$ and modal matrix $\Phi$.\footnote{In this section, the $r$th vector of the modal matrix (commonly known as the $r$th modeshape) will be denoted by $\B{\phi}^{(r)}$.}

Experimental modal analysis (EMA) \cite{ewins2009modal} is a general term for methods that attempt to specify the modal properties from measured structural dynamics data. Broadly, these methods can be divided into methods that operate in the time and frequency domains. In this paper, two EMA methods are presented on the proposed Hawk dataset. In the frequency domain, the method of rational fractional polynomials (RFP) \cite{Richardson1985} is employed. In the time domain, an output-only stochastic subspace identification (SSI) is shown.

\subsection{Modal analysis: Frequency domain}

Frequency domain methods for EMA rely on approximating the form of the frequency response function (FRF), often employing techniques from curve fitting. Assuming that the dynamics are linear, the FRF is a linear transfer function from the Fourier-transformed input force  (N) to the output acceleration (g) that has a generalised form that depends only on the modal properties. In order to keep the notation in this section general (and in keeping with the system identification literature), the input forcing spectra will be denoted $X(\B{\omega})$ and the output acceleration $Y(\B{\omega})$, where $\B{\omega}$ is the vector of frequencies for which data were recorded. The accelerance FRF can be expressed as,

\begin{equation}
	H(\B{\omega}) = \frac{Y(\B{\omega})}{X(\B{\omega})}
\end{equation}

The expression for a general accelerance FRF in terms of the modal parameters is,

\begin{equation}
	H_{il}(\B{\omega}) = \sum_{r=1}^n \frac{-\B{\omega}^2\B{\phi}_i^{(r)}\B{\phi}_j^{(r)}}{-\B{\omega}^2 + 2j\zeta_r\omega_r\B{\omega} +\omega_r^2}
	\label{eqn:frf_gen}
\end{equation}

Where $\omega_r$, $\zeta_r$ and $\phi^{(r)}$ are the natural frequency, damping ratio and modeshape corresponding to the $r$th vibration mode and $i$ and $l$ are the indices of the input and output location respectively.

The method of RFP works by approximating \eqref{eqn:frf_gen} as a ratio of two polynomials. Considering here a single-input, single output (SISO) system comprised of $n$ modes, one has,

\begin{equation}
	H(\B{\omega}) = \frac{\sum_{k=0}^m a_k (j\B{\omega})^k}{\sum_{k=0}^{n} b_k (j\B{\omega})^k}
 \label{eqn:rfp_poly}
\end{equation}

where the coefficients $a_k$ and $b_k$ depend on the modal properties. Note that the number of terms in the numerator polynomial does not necessarily have to be the same as the denominator. Additional terms in the numerator polynomial can be used to compensate for out-of-band effects (i.e modes from outside the frequency range of interest).

RFP learns the $a_k$ and $b_k$ coefficients from the measured $H$ and then infers the modal properties. Rather than learn the coefficients by optimisation, it is preferable to construct a least squares scheme. In order for the coefficients to be identifiable we must include a constraint that fixes one of the polynomial coefficients (or else there would be a constant factor invariance). Here, after \cite{Richardson1985}, the lead coefficient of the denominator is arbitrarily set to unity $b_n=1$.  In order to simplify notation, the vectors $\B{a}, \B{b}$ are defined as the unknown coefficients of the numerator (up to order $m$) and the denominator (up to order $n-1$) polynomials respectively.

Representing the rational FRF form above in matrix form,

\begin{equation}
	H = (\Phi_{\B{a}} \B{a}) (b_n (j\B{\omega})^n + \Phi_{\B{b}} \B{b})^{-1}
\end{equation}

where the matrices $\Phi_{\B{a}}$ and $\Phi_{\B{b}}$ are polynomial bases of order $n$ and $m$ respectively for the vector of frequency lines $\B{\omega}$, and $(j\B{\omega})^n$ is the vector $j\B{\omega}$ raised to the power $n$ element-wise. In order to construct the least squares scheme, the error between the measured $\hat{H}(\B{\omega})$ and predicted FRFs can be expressed as,

\begin{equation}
	\B{e} = \Phi_{\B{a}}\B{a} - ((j\B{\omega})^n + \B{b} \Phi_{\B{b}})\hat{H}(\B{\omega})
\end{equation}

In order to simplify the forthcoming notation (and for compatibility with the notation of \cite{Richardson1985}), the following notation is introduced here,

\begin{equation}
	P = \Phi_{\B{a}}
\end{equation}

\begin{equation}
	T = \Phi_{\B{b}} \hat{H}(\B{\omega})
\end{equation}

\begin{equation}
	\B{w} = b_n (j\B{\omega})^n	\hat{H}(\B{\omega})
\end{equation}

Thus the objective function becomes,

\begin{equation}
	J(\B{a},\B{b} | \hat{H}) = \B{e}^\dag \B{e} = (P\B{a} - (\B{w} + T\B{b}))^\dag (P\B{a} - (\B{w} + T\B{b}))
\end{equation}

where $\dag$ represents the Hermitian transpose. The above is a positive quadratic in both $\B{a}$ and $\B{b}$ and therefore has a single optimum ($\B{a}^*$, $\B{b}^*$), obtained exactly by setting the partial derivatives in $\B{a}$ and $\B{b}$ to zero,

\begin{equation}
	\frac{\partial J(\bm{a}, \bm{b} | \hat{H})}{\partial \bm{a}} \bigg|_{\bm{a}^*,\  \bm{b}^*}=0
\end{equation}

and,

\begin{equation}
	\frac{\partial J(\bm{a}, \bm{b} | \hat{H})}{\partial \bm{b}} \bigg|_{\bm{a}^*,\  \bm{b}^*}=0
\end{equation}

Now evaluating these derivatives starting with the former condition,

\begin{equation}
	\frac{\partial J}{\partial \bm{a}} =
	2\mathbb{Re}(P^\dag P) \bm{a}
	- 2\mathbb{Re}(P^\dag T) \bm{b}
	- 2\mathbb{Re}(P^\dag \bm{w})
\end{equation}

where $\mathbb{Re}$ denotes the real part of the complex vector. Next, for the second condition,

\begin{equation}
	\frac{\partial J}{\partial \bm{b}} =
	- 2\mathbb{Re}(T^\dag P) \bm{a}
	+ 2\mathbb{Re}(T^\dag T) \bm{b}
	+ 2\mathbb{Re}(T^\dag \bm{w})
\end{equation}

Finally, applying the condition of zero gradient at the optimum, one has the block-linear system that can be solved for the polynomial coefficients.

\begin{equation}
	\mathbb{Re}\left(\begin{bmatrix} P^\dag P & -P^\dag T \\ - T^\dag P & T^\dag T \end{bmatrix}\right) \begin{bmatrix}\bm{a}\\ \bm{b}\end{bmatrix} = \mathbb{Re}\left(\begin{bmatrix} P^\dag \bm{w} \\ -T^\dag \bm{w} \end{bmatrix}\right)
\end{equation}

It is worth noting that this expression differs in several respects from the expression in \cite{Richardson1985} due to the presence of several errors in the original manuscript. Finally the modal properties (natural frequencies and damping ratios) can be recovered from the coefficients of the denominator polynomial as,

\begin{equation}
	\omega_r = |R_r|
\end{equation}

\begin{equation}
	\zeta_r = \frac{-\text{Re}(R_r)}{|R_r|}
\end{equation}

where $R_r$ is the $r$ th complex-conjugate root of the denominator polynomial in \eqref{eqn:rfp_poly}. Extraction of the modeshapes is also possible, requiring both the coefficients of the numerator and the denominator polynomials. However it is not considered here.

\begin{figure}
	\centering
	\includegraphics[width=\fw\linewidth]{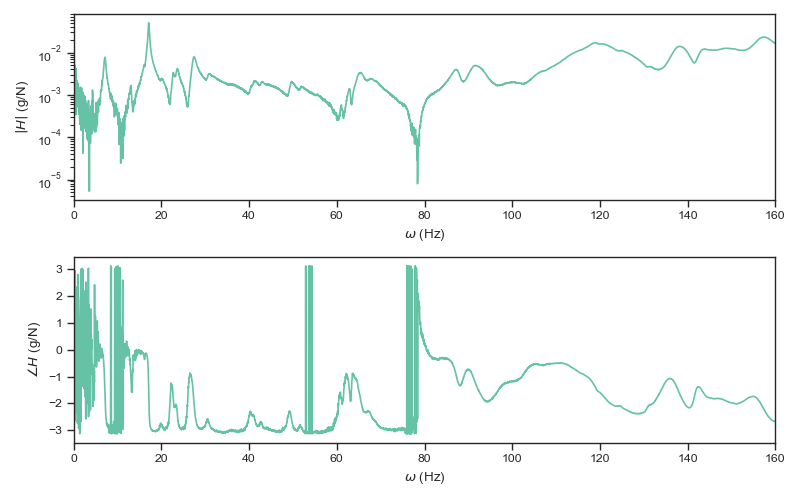}
	\caption{FRF from sensor LTC-05 subject to burst-random excitation at 0.4V.}
	\label{fig:frf}
\end{figure}

In Figure \ref{fig:frf}, the FRF of the hawk at sensor LTC-05 (Lower trailing edge centre, burst-random excitation, 0.4V LMS input voltage, averaged over 10 repeats) is plotted. In order to demonstrate the RFP approach, some modal properties will be extracted from this signal.

\begin{figure}
	\centering
	\includegraphics[width=\fw\linewidth]{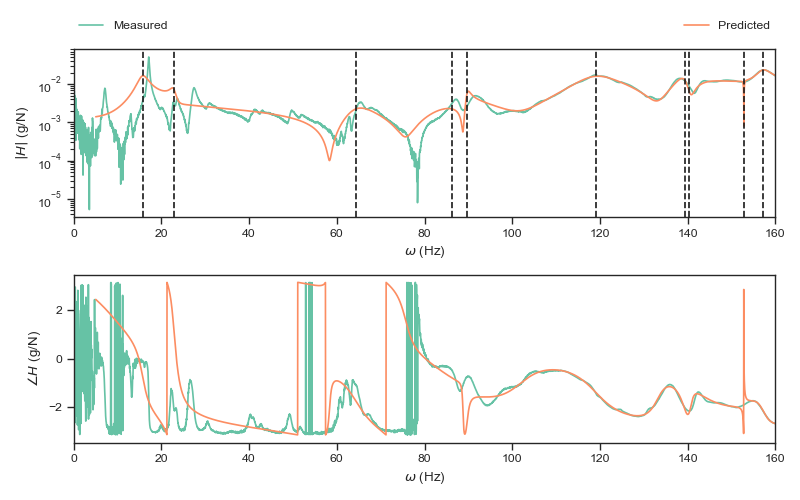}
	\caption{Measured vs. identified RFP FRF for the sensor at LTC-05, vertical lines are the identified natural frequencies.}
	\label{fig:rfp1}
\end{figure}

As a first analysis, all frequency lines are included in the computation of the modal properties with the number of modes set to 10 (corresponding to a 20th order polynomial). The resulting FRF is plotted in Figure \ref{fig:rfp1} with dashed lines corresponding to the natural frequencies. As can be seen, the higher frequencies are well approximated but there is significant errors in the lower frequencies and many of the dominant modes have been missed entirely. This is due to an inherent bias toward higher frequencies in the RFP method that can be attributed to the choice of the error criterion $\B{e}$.\footnote{This bias can be seen by noting that the choice of $\B{e}$ does not correspond directly to the difference between the measured and predicted FRF, but a difference that is scaled by the frequency.}

\begin{figure}
	\centering
	\includegraphics[width=\fw\linewidth]{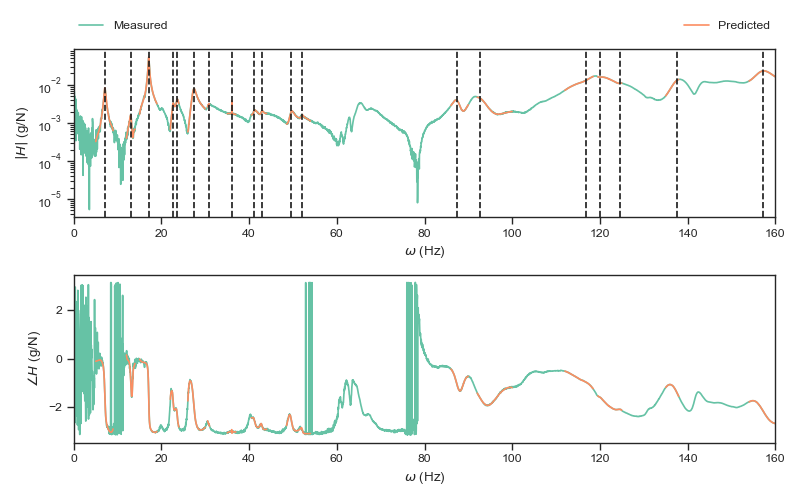}
	\caption{Measured vs. identified banded RFP FRF for the sensor at LTC-05, vertical lines are the identified natural frequencies.}
	\label{fig:rfp2}
\end{figure}

A more practical approach (indicative of how frequency-domain RFP is often implemented) is to instead perform the curve fit on a number of pre-selected bounds. In order to compensate for the effects of `out-of-bound' modes, additional terms can be added to the numerator polynomial. Here, for each range, 6 additional terms are added to the numerator polynomial. The predicted FRF and natural frequencies are plotted in Figure \ref{fig:rfp2} for identification bounds given in Figure \ref{tbl:bnds}. In the figure, the quality of the FRF fit is visibly superior to the global fit and the the peak locations are well identified within the selected bands. Note that the total FRF can be reconstructed by first estimating the damping ratios and then using the general form of the modal transfer function although this is not considered here.

It is clear that such an approach cannot be considered to be `automatic' and that the performance comes to depend strongly on the choice of the identification ranges. However it is important to consider such approaches as a baseline to which more advanced methods might be compared.

\begin{table}
	\centering
	\caption{Bands and identified modal parameters from the banded RFP identification.}
	\label{tbl:bnds}
	\begin{tabular}{lllll}
		Lower bound (Hz) & Upper bound (Hz) & Modes in band & $\omega_n$  (Hz) & $\zeta_n$          \\\hline
		5                & 9                & 1             & 7.094            & 0.02434            \\
		12               & 14               & 1             & 13.05            & 0.01552            \\
		15               & 19               & 1             & 17.1             & 0.005918           \\
		22               & 24               & 2             & 22.66, 23.56     & 0.008452, 0.009352 \\
		26               & 30               & 1             & 27.35            & 0.01584            \\
		30.5             & 31               & 1             & 30.77            & 0.0004578          \\
		35               & 37               & 1             & 36.06            & 0.00005626         \\
		40.5             & 44               & 2             & 41.16, 42.85     & 0.008827, 0.007161 \\
		48.5             & 54               & 2             & 49.61, 52.07     & 0.00995, 0.00469   \\
		86               & 90               & 1             & 87.41            & 0.01138            \\
		92               & 100              & 1             & 92.54            & 0.02798            \\
		112              & 118.5            & 1             & 116.9            & 0.01307            \\
		119.5            & 122              & 1             & 120              & 0.003427           \\
		122              & 125              & 1             & 124.6            & 0.008256           \\
		135              & 138              & 1             & 124.6            & 0.008256           \\
		154              & 162              & 1             & 157.1            & 0.01505            \\
	\end{tabular}
\end{table}

\subsection{Modal analysis: Time domain}

Several approaches to EMA consider the identification problem in the time-domain. Of these, a common approach is stochastic subspace identification (SSI) \cite{VanOverschee1994}. The SSI approach outlined here has the added advantage that it operates in an output-only setting (subject to a broadband excitation assumption). This is especially attractive for applications whereby measurement of system input in-service (such as in an aircraft during flight) is challenging.

The collection of time-series data from the Hawk allows the effectiveness of SSI to be tested. Here, a classical approach of ``covariance-driven'' SSI (Cov-SSI) is used. In this work the terms SSI and Cov-SSI will now be used interchangeably. A full background to SSI will not be presented in this paper in the name of brevity, instead the reader is referred to \cite{VanOverschee1994, Katayama2005} for general introductions to the family of methods or towards \cite{peeters2001stochastic} in reference to the output-only \emph{operational modal analysis} (OMA) setting or more recently with respect to uncertainty quantification \cite{reynders2016uncertainty} (although the subject of uncertainty is not addressed in this work, see \cref{sec:challenges}).

Briefly, the application of SSI requires formulation of the (block) Hankel matrices of the ``past'' and ``future''. These are constructed through stacking time-delayed (lagged) versions of the measured time-series responses of the system. The number of lags chosen in the Hankel matrices defines the maximum order of the system which can be identified, where this maximum order is twice the maximum number of modes. After constructing these matrices, $Y_p$ and $Y_f$, for the past and future respectively, the cross covariance can be computed for $N$ datapoints as,

\begin{equation}
    \Sigma = \begin{bmatrix}
        \Sigma_{pp} & \Sigma_{pf} \\
        \Sigma_{fp} & \Sigma_{ff} \\
    \end{bmatrix} = \frac{1}{N}\begin{bmatrix}
        Y_p \\ Y_f
    \end{bmatrix}\begin{bmatrix}
        Y_p\tran & Y_f\tran
    \end{bmatrix}
\end{equation}

This cross covariance is partitioned to consider the auto-covariance within the $Y_p$ and $Y_f$, and the cross covariance between them. To perform SSI, a canonical correlation analysis between these two sets of variables is performed based on the covariances. Mathematically, a singular value decomposition (SVD) is taken of the normalised covariance, $$ USV\tran = \Sigma_{pp}^{1/2}\Sigma_{pf}\Sigma_{ff}^{1/2}$$ 

The singular values (along the diagonal of $S$) are sorted in descending order and, to select an appropriate model order, a number of the smallest of these values is discarded. If one wishes to construct the consistency (stabilisation) diagram, sequentially more and more values are discarded such that the model order is reducing and identification is performed at each of these orders. Since the computationally intensive stage of SSI is the formation of the Hankel matrices, covariances and SVD, this procedure for generating the consistency diagram is quite efficient. The truncated set of singular values can be denoted as $\hat{S}$, the associated columns of $U$ and $V$ are also removed with these truncated matrices as $\hat{U}$, $\hat{V}$. Given that it can be shown that the cross covariance is the product of the observability and controllability matrices, $\Sigma_{fp} = \mathcal{O}\mathcal{C}$, these can be recovered by: 

\begin{subequations}
    \begin{equation}
        \mathcal{O} = \Sigma_{ff}^{1/2}\hat{U}\hat{S}^{1/2}
    \end{equation}
    \begin{equation}
        \mathcal{C}\tran = \Sigma_{pp}^{1/2}\hat{V}\hat{S}^{1/2}
    \end{equation}
\end{subequations}

From $\mathcal{O}$ the system matrices $A$ and $C$ of a linear dynamic state space model with states $\vec{x}$:

\begin{equation}
    \begin{aligned}
        \vec{x}_{t+1} = A\vec{x}_t + \vec{v}_t\\
        \vec{y}_t = C\vec{x}_t + \vec{w}_t
    \end{aligned}
\end{equation}

\noindent where $\vec{v}_t$ and $\vec{w}_t$ are samples from multivariate Gaussian distributions with covariances $Q$ and $R$ respectively. %
Given that the observability matrix is constructed as, $$\mathcal{O} = \begin{bmatrix}
    C \\ CA \\ CA^2 \\ \vdots \\ CA^K
\end{bmatrix}$$ where $K$ is the maximum model order considered, the $C$ matrix of the system can be recovered as the first block row of this matrix and then the $A$ matrix can be recovered via a least squares solution. %
The $A$ and $C$ matricies fully define the dynamics of the system and it is therefore possible to recover the modal properties. %
A common step in the SSI approach is to vary the size of the $A$ and $C$ matrices of the system and hence the model order by reducing the level of truncation of the singular values (usually in steps of two since the underlying system is assumed to be of a set of second order modes). Then, comparing the recovered modal properties a  consistency\footnote{Here the term \emph{consistency} is preferred to  \emph{stabilisation} since there is not any link between the pole stability in terms of a requirement for the system to possess positive damping and the similarity of the modal properties between each model order. Therefore, the terminology could be ambiguous and give some implication about the system which may be misinterpreted.} (also ``stabilisation'') diagram is constructed based on the changes in these modal properties as the model order varies.

\begin{figure}
	\centering
	\includegraphics[width=\fw\linewidth]{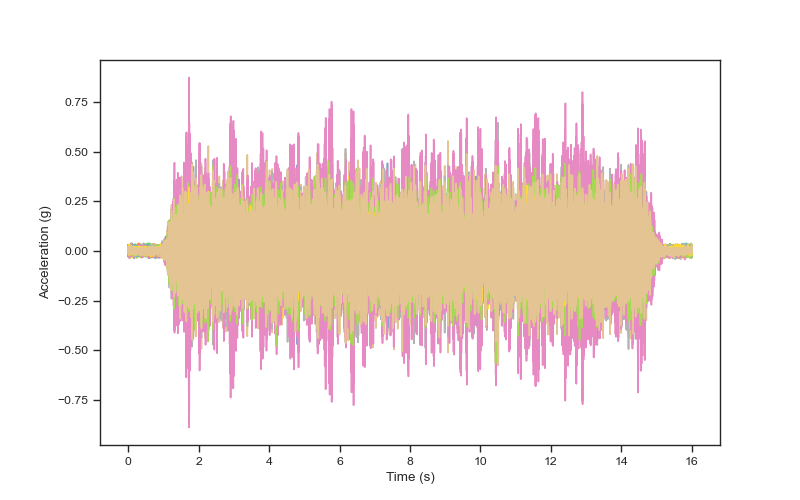}
	\caption{Time series collected at each of the 55 sensor locations subject to closed-loop random phase excitation at 0.4V.}
	\label{fig:NI}
\end{figure}

In Figure \ref{fig:NI}, the time series from all accelerometers in the random phase amplitude test (excitation level 0.4V) are plotted. Each burst signal represents 32,768 datapoints captured over a 16s second interval. Unlike the RFP approach above, SSI is able to simultaneously incorporate data from all of the sensors at once. In order to speed up the computation of the SSI, the time series are decimated by a factor of 5 prior to the calculation of the Hankel matrices.

\begin{figure}
	\centering
	\includegraphics[width=\fw\linewidth]{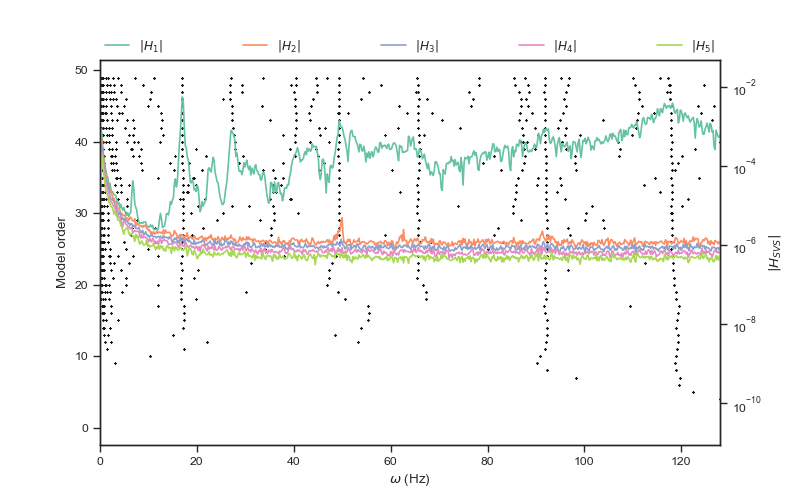}
	\caption{SSI consistency diagram (black dots) compared to the first 5 components of the SVS (lines), subject to closed-loop random phase excitation at 0.4V.}
	\label{fig:SSI}
\end{figure}

In order to evaluate the performance of the SSI, the singular-valued spectrum (SVS) \cite{Hassani2007} is calculated. It is useful to consider the SVS as a convenient visualisation as it is expected that all modes should be evident in only the first few singular values, whereas any given sensor may not have all modes visible (for example if a sensor is close to a nodal point in one of the modeshapes). In Figure \ref{fig:SSI}, the modal consistency diagram is plotted over the SVS. This diagram plots the predicted natural frequencies at each model order up to a maximum of 50. A standard approach in EMA is to interpret the `true' modes as those that are `consistent' in the diagram (that is they are present at each model order).

As can be seen in the figure, the SSI approach has identified the major resonances well although there are a significant number of non-physical modes especially in the higher order-models. Once again it useful to interpret this result as a baseline against which more advanced output-only time domain methods might be considered.

\section{Structural health monitoring}
\label{sec:shm}

To equip engineering asset management teams with the information to make more optimal operation, maintenance and repair decisions for their high value infrastructure, the last several decades have seen a rise in developments of structural health monitoring strategies to provide insight into the condition of structures using remote sensing \cite{farrar2012structural}. A health monitoring system is defined as the implementation of a strategy that facilities the real time assessment of the condition of a structure - often for civil infrastructure or high value assets such as aircraft - for ideally the span of the structure's operational life. Broadly speaking, installing an SHM strategy first involves acquiring data relating to the dynamic behaviour of the structure, such as accelerometer measurements. Features sensitive to damage are then extracted, such as the natural frequencies of the structure, which are subsequently used as inputs to a statistical model that relate the features to some information pertaining to the damage state of the system. 

The necessary complexity and sophistication of the model used can vary significantly, and is generally dependent on which level of the SHM hierarchy one is interested, of which there are five stages \cite{worden2004overview}:

\begin{enumerate}
        \item Detection - is damage present?
        \item Localisation - where is the damage?
        \item Classification - what type of damage?
        \item Assessment - how severe is the damage?
        \item Prognosis - what is the remaining useful life of the structure?
\end{enumerate}

As progress is made up the hierarchy, the complexity of the task increases, with classification, assessment and prognosis being particularly challenging. Regardless of the complexity of the SHM task that is being attempted, to transition monitoring algorithms from research to operational implementation, it is vital that the algorithms are validated on real world data to ensure that they are robust and reliable such that the risk of malfunction is inline with the guidelines of the relevant industry. Datasets that can be used to validate such methods are therefore critical in progressing structural health monitoring into engineering practice.  

In the following section, an overview of some basic SHM approaches that have been applied to the dataset introduced in this paper will be provided, including the use of both unsupervised and supervised machine learning techniques. In the analysis that follows, data are taken from the damage simulation tests in the frequency domain. In total, 27 test runs are taken at amplitude levels between 0.4V and 2.0v for three levels of damage severity (added mass) and at three different locations along the wing. For undamaged condition data, the tests are taken from the burst-random amplitude ramp test series with excitation values between 0.4V and 2.0v. In total this corresponds to 32 test series and 320 FRFs at each sensor.

In this paper, two SHM approaches are presented. As a supervised learning benchmark, a Gaussian mixture model (GMM) is fitted to the principal components of the natural frequencies. In the unsupervised case, a novelty detector based on the Mahalanobis distance is employed with damage sensitive spectral lines used as features.

\subsection{Damage detection: Supervised learning}

\begin{figure}
	\centering
	\includegraphics[width=\fw\linewidth]{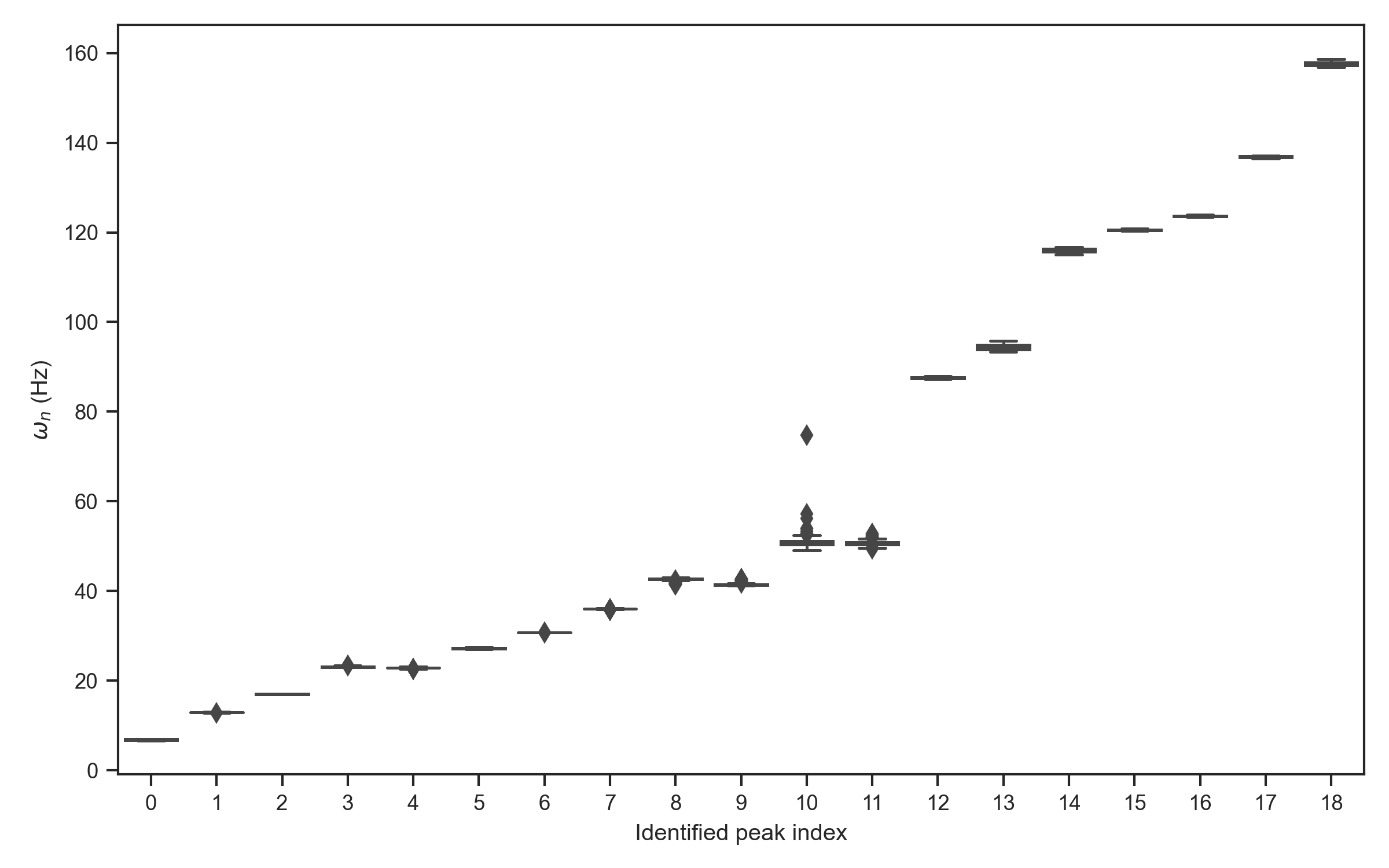}
	\caption{Boxplot of the natural frequencies from the RFP analysis for each of the testing scenarios.}
	\label{fig:box}
\end{figure}

In order to demonstrate a first analysis in supervised SHM, a GMM is fitted to features arising from a modal identification as per the previous section. Natural frequencies are extracted using an RFP approach with ranges defined in \Cref{tbl:bnds}. Overall, 18 natural frequencies are derived for each sensor for each of the test runs. In order to reduce the variance in the identification process, the natural frequencies are averaged over each sensor location. \Cref{fig:box} plots a boxplot of the resulting features. As can be seen in the figure, there are several outliers in the identification of the 10th peak, indicating that this peak has not been well identified. To avoid the variance in this mode dominating the features, the 10th natural frequency is removed from the feature set for an overall feature dimension of 18.

\begin{figure}
	\centering
	\includegraphics[width=\fw\linewidth]{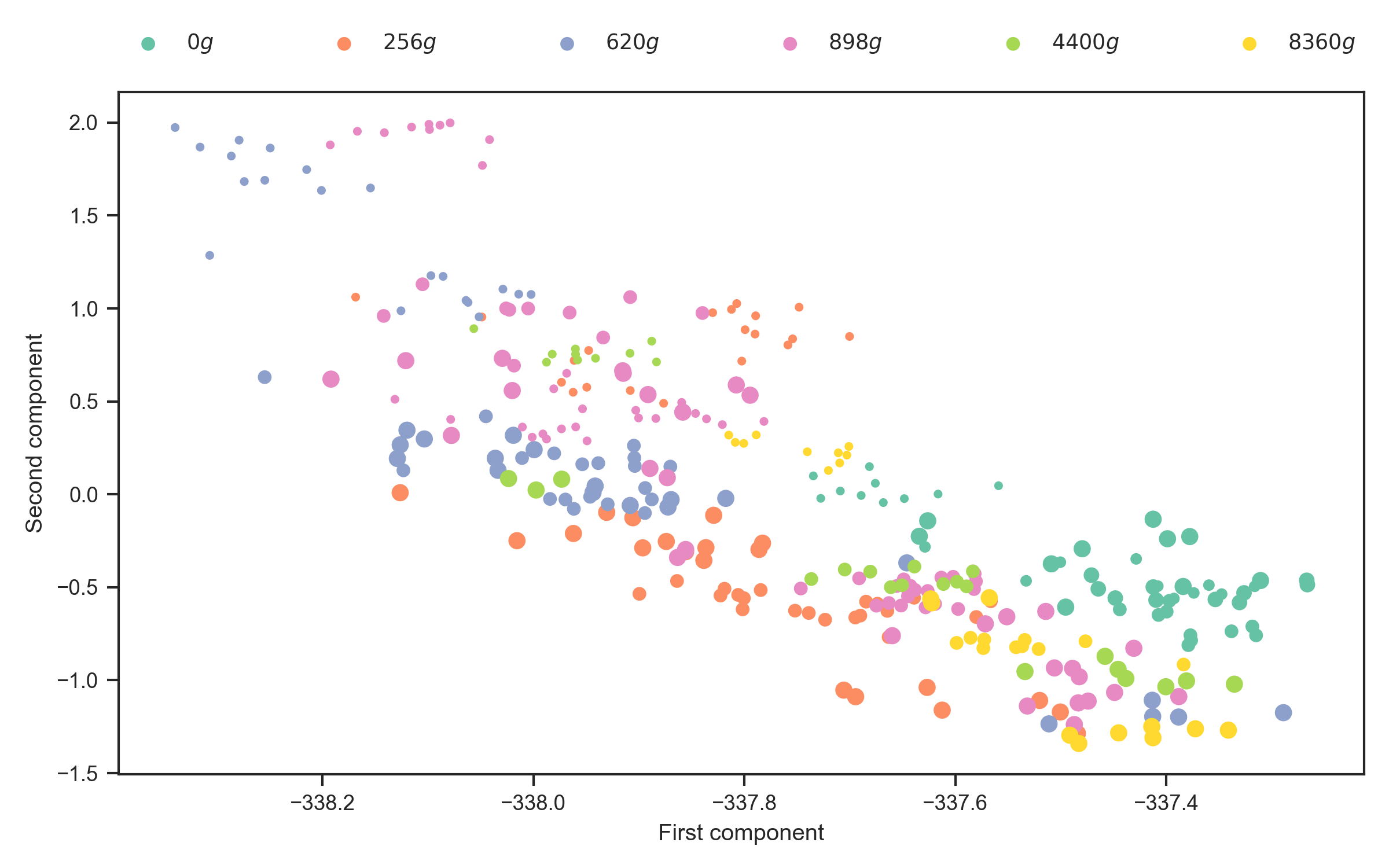}
	\caption{First two principal components of the natural frequency features. In the plot colour denotes damage severity (added mass) and marker size denotes excitation level (larger markers for higher excitation levels).}
	\label{fig:PCA}
\end{figure}

In order to further reduce the feature dimension, a principal component analysis (PCA) decomposition is taken on the natural frequencies. The first two principal components of the resulting features are plotted in Figure \ref{fig:PCA}. In the plot the samples are coloured by damage extent and the relative sizes of the markers represent the excitation level. Considering the data in Figure \ref{fig:PCA}, it is clear that the classes are not linearly separable and therefore the GMM classifier might not be expected to perform well. However it is still a useful result to consider the GMM as baseline for clustering performance such that advanced approaches might be compared to first analyses.

\begin{figure}
	\centering
	\begin{subfigure}[b]{0.49\textwidth}
		\centering
		\includegraphics[width=\textwidth]{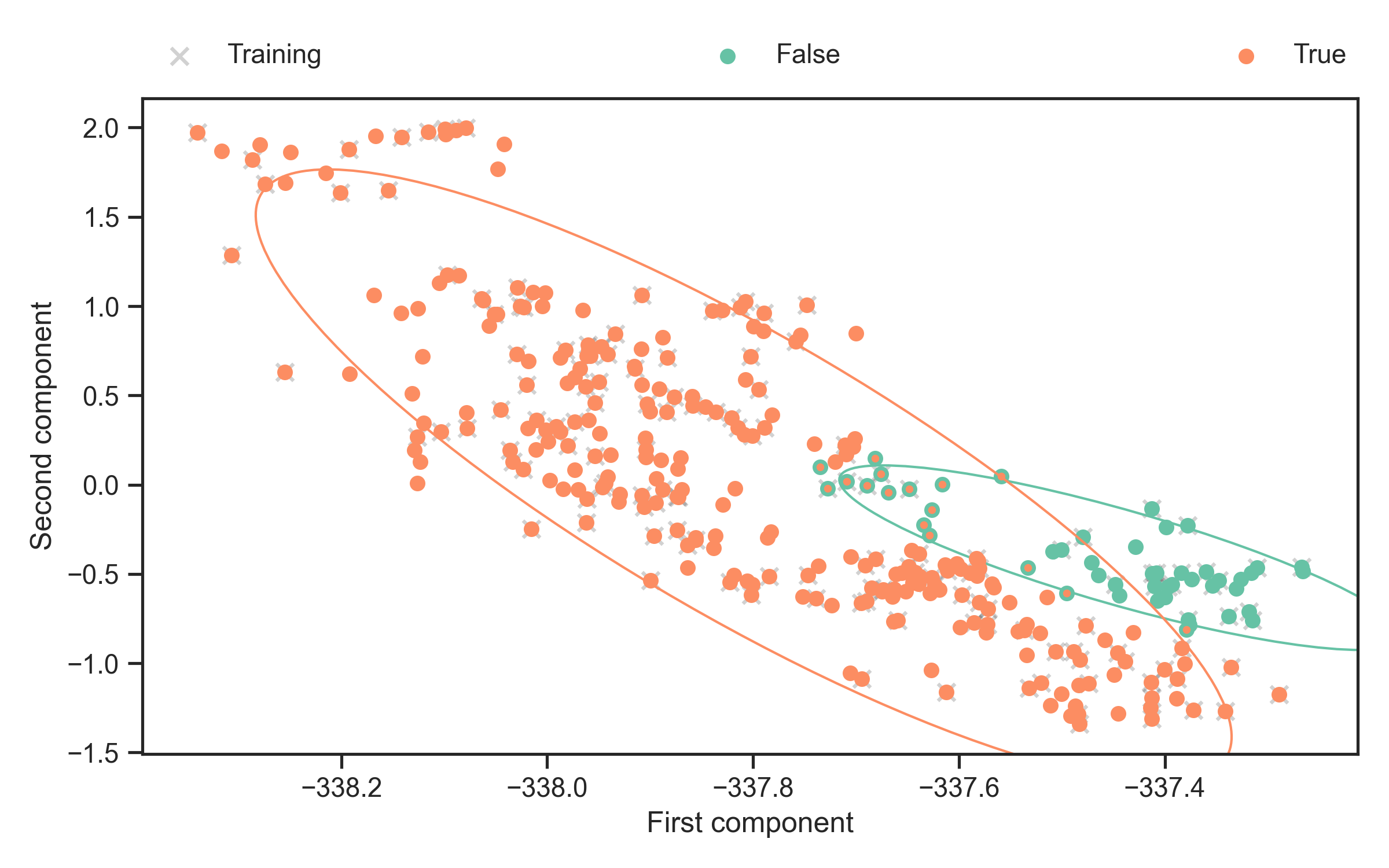}
		\caption{Presence of damage.}
		\label{fig:gmm_11}
	\end{subfigure}
	\hfill
	\begin{subfigure}[b]{0.49\textwidth}
		\centering
		\includegraphics[width=\textwidth]{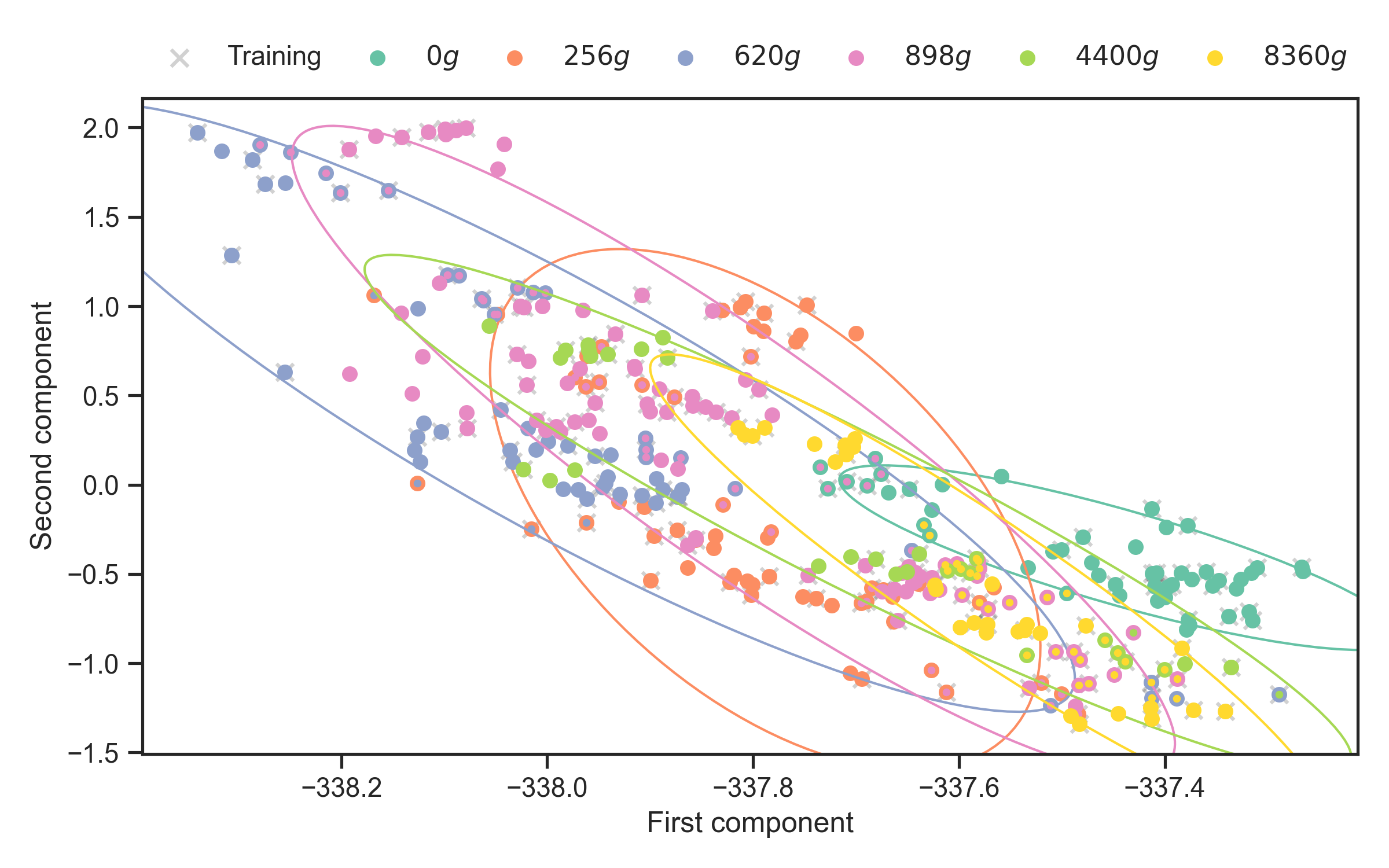}
		\caption{Added mass amount.}
		\label{fig:gmm_12}
	\end{subfigure}
	\hfill
	\begin{subfigure}[b]{0.49\textwidth}
		\centering
		\includegraphics[width=\textwidth]{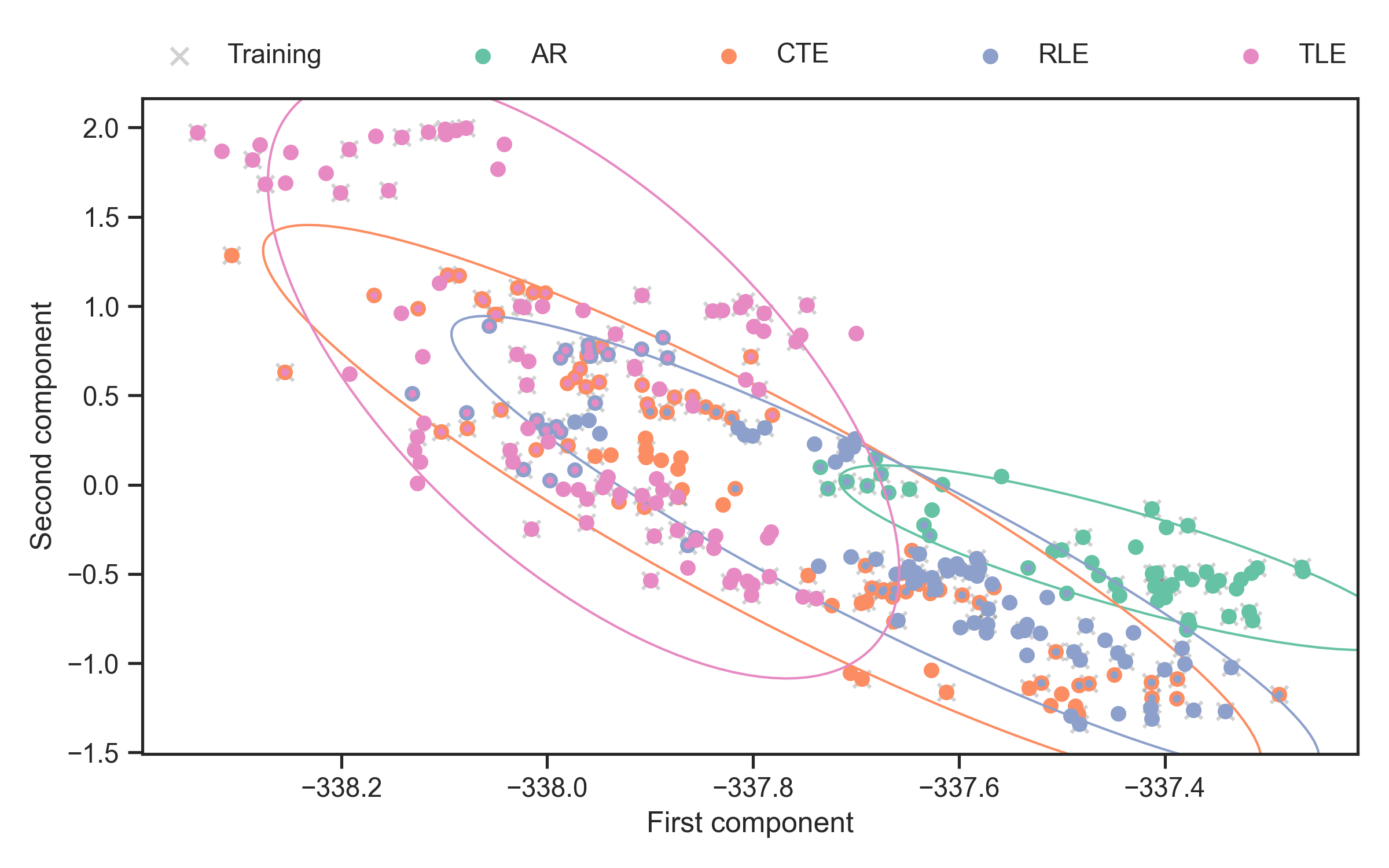}
		\caption{Damage location.}
		\label{fig:gmm_21}
	\end{subfigure}
	\hfill
	\begin{subfigure}[b]{0.49\textwidth}
		\centering
		\includegraphics[width=\textwidth]{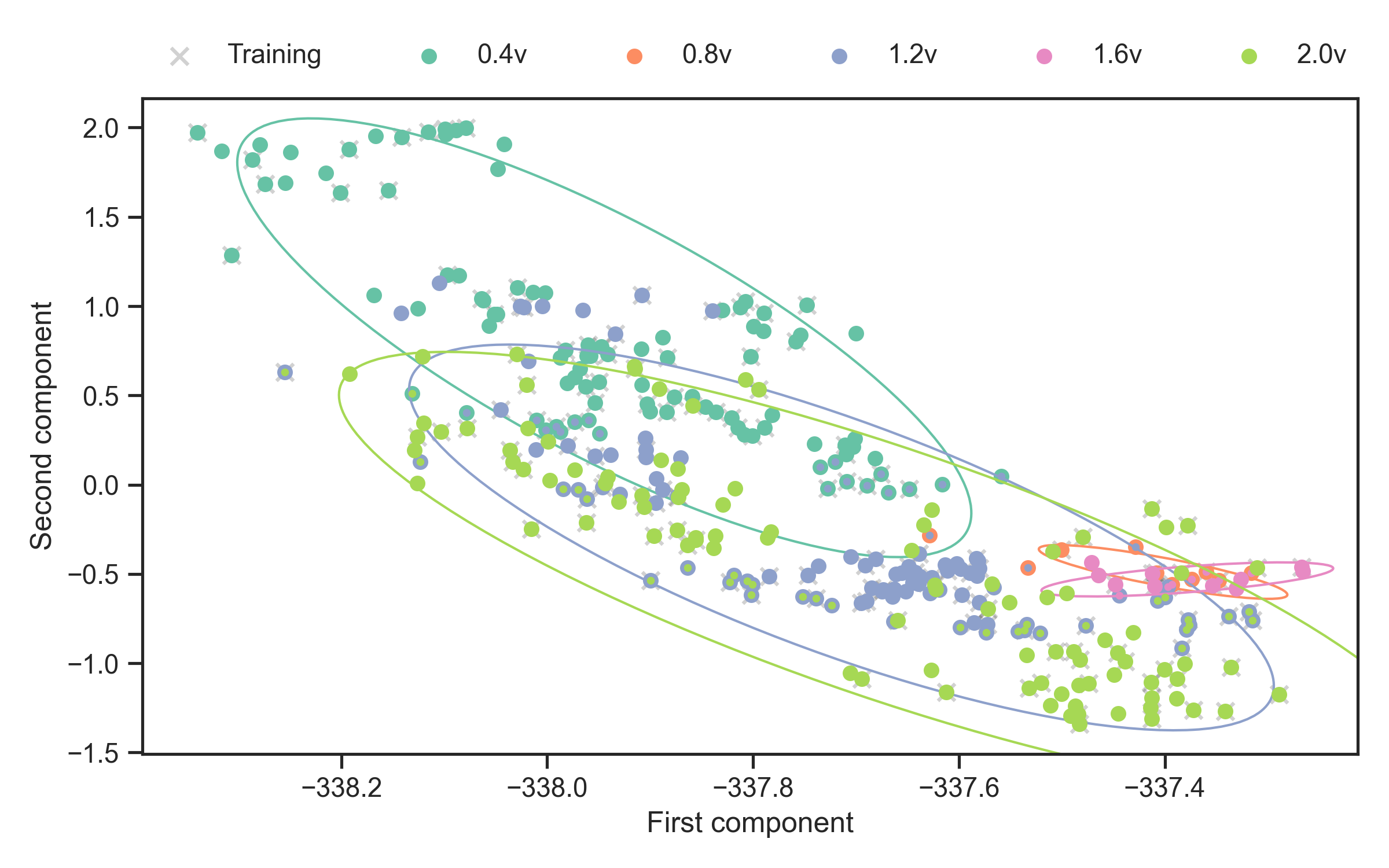}
		\caption{Excitation level.}
		\label{fig:gmm_22}
	\end{subfigure}
	\caption{Supervised GMM clustering results on the first two principal components of the natural frequency features.}
	\label{fig:gmm}
\end{figure}

\begin{figure}
	\centering
	\begin{subfigure}[b]{0.49\textwidth}
		\centering
		\includegraphics[width=\textwidth]{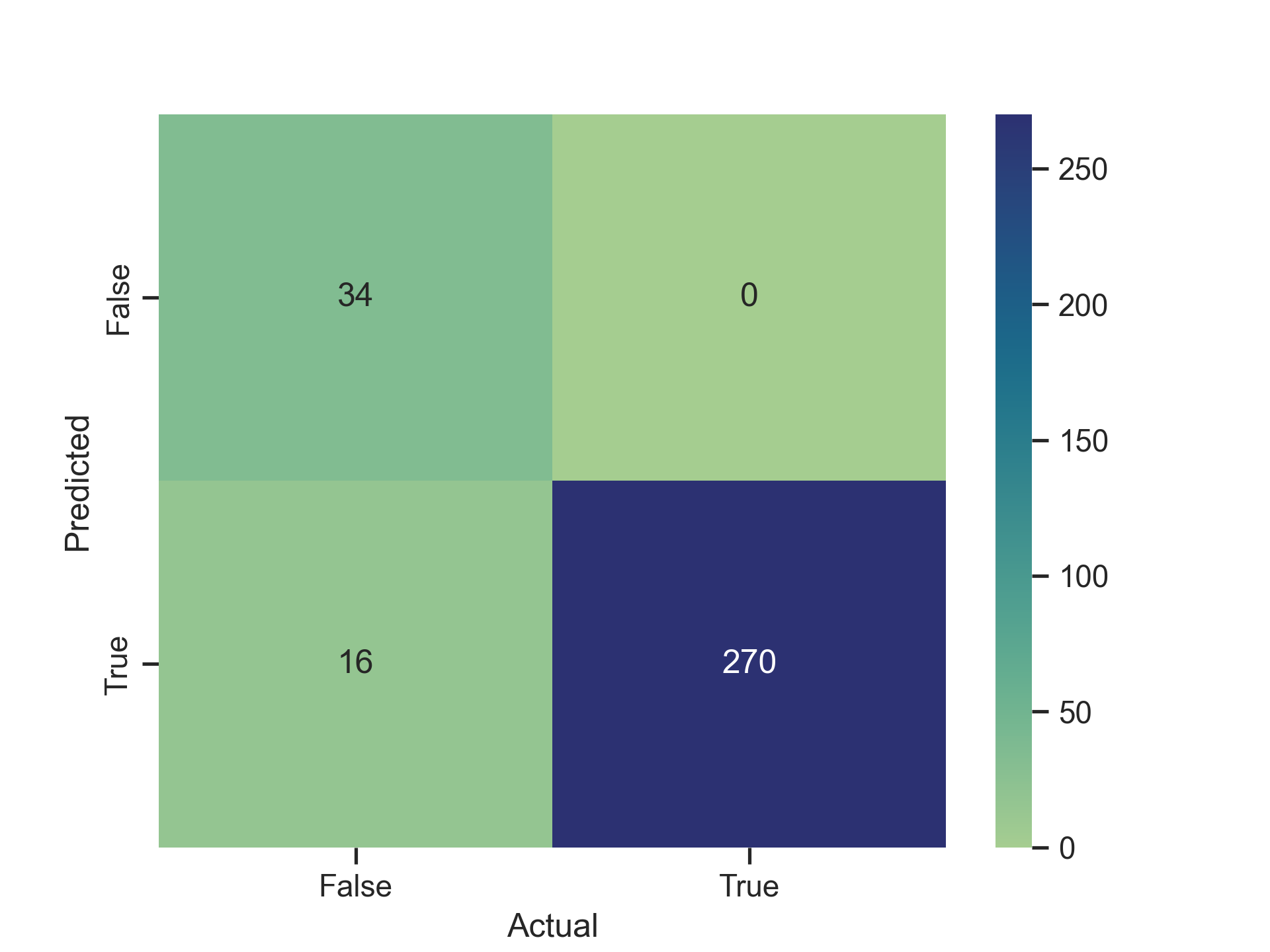}
		\caption{Presence of damage.}
		\label{fig:con_11}
	\end{subfigure}
	\hfill
	\begin{subfigure}[b]{0.49\textwidth}
		\centering
		\includegraphics[width=\textwidth]{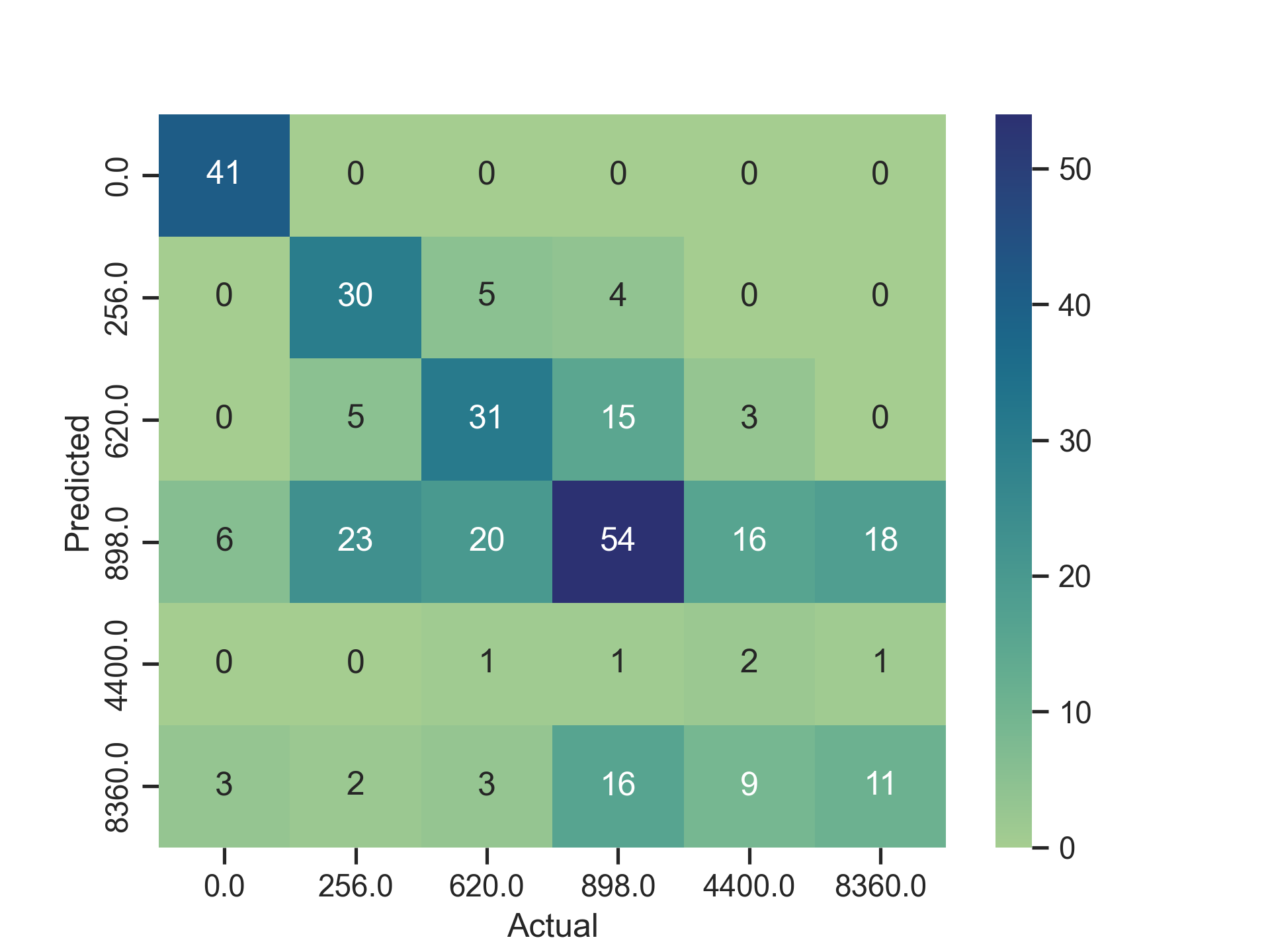}
		\caption{Added mass amount.}
		\label{fig:con_12}
	\end{subfigure}
	\hfill
	\begin{subfigure}[b]{0.49\textwidth}
		\centering
		\includegraphics[width=\textwidth]{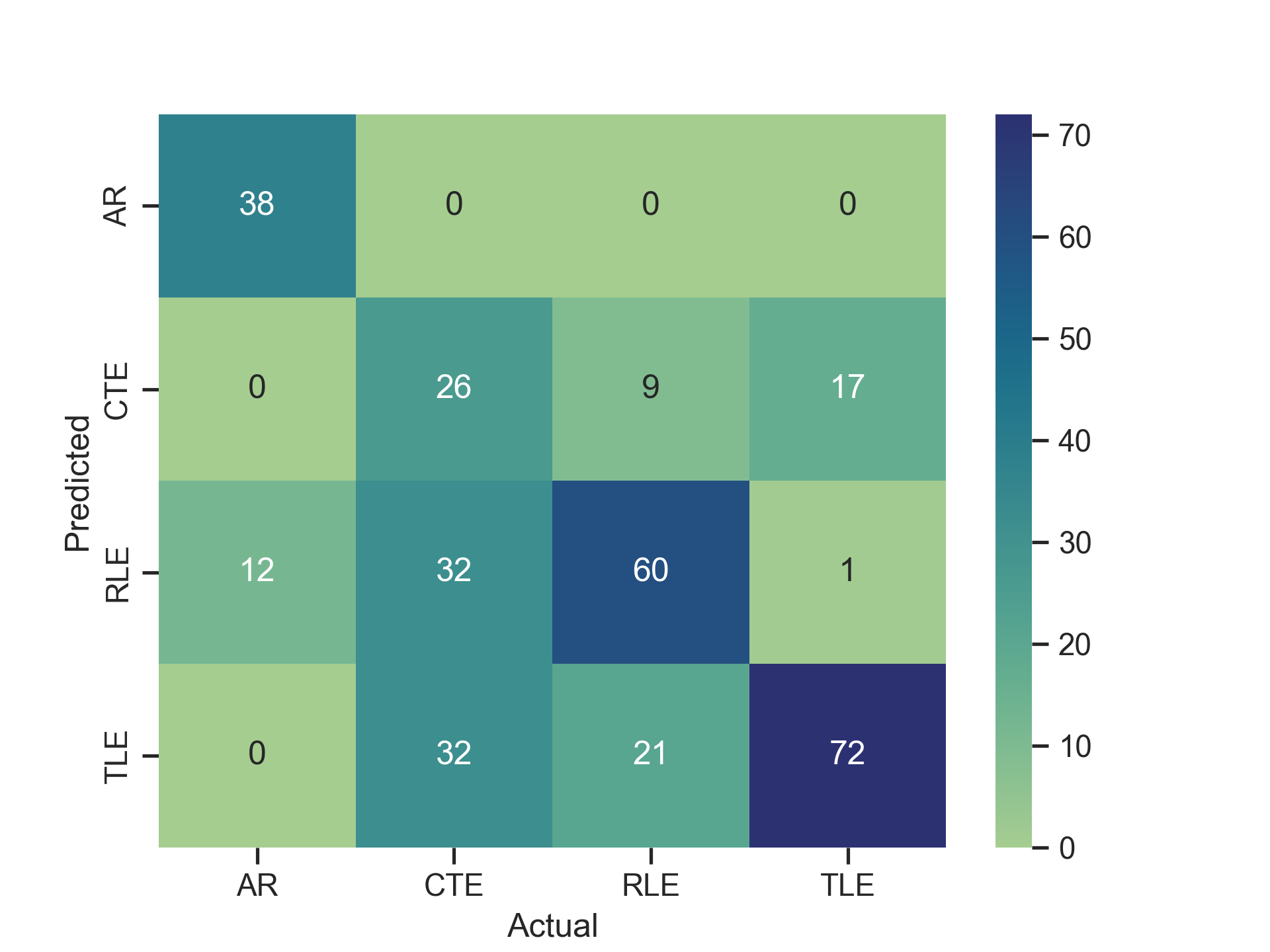}
		\caption{Damage location.}
		\label{fig:con_21}
	\end{subfigure}
	\hfill
	\begin{subfigure}[b]{0.49\textwidth}
		\centering
		\includegraphics[width=\textwidth]{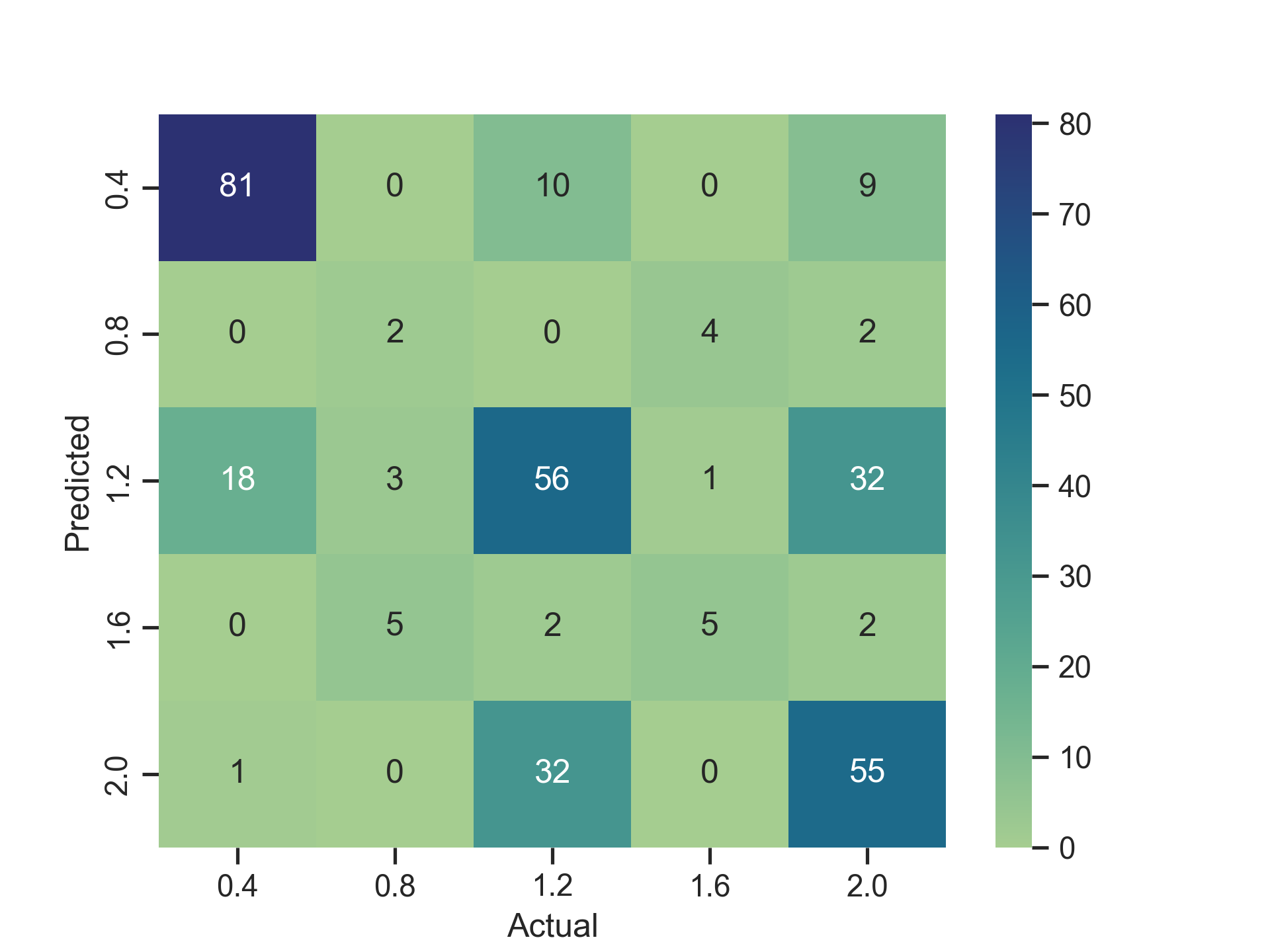}
		\caption{Excitation level.}
		\label{fig:con_22}
	\end{subfigure}
	\caption{Confusion matrices from the supervised GMM clustering.}
	\label{fig:con}
\end{figure}

In order to classify the data, a GMM is trained. For training data, a subset of half of the training examples is used with balanced membership of each class in the training and testing sets. The training data are then used to empirically fit Gaussian distributions to each of the classes by evaluating the sample mean and covariances. The overall probability that an unseen example $x^*$ has a label $y^*$ corresponding to the $k$th class can be written,

\begin{equation}
	p(y^* | X) = \sum_k \pi_k \mathcal{N}(\bm{\mu}_k, \bm{\Sigma}_k)
\end{equation}

where $X$ are the training data and,

\begin{equation}
	\pi_k = \frac{n_k}{N}
\end{equation}

the proportion of training samples in the $k$th class and,

\begin{equation}
	\bm{\mu}_k = \mathbb{E}[\bm{x}_i] \quad \forall i \ \ \text{s.t.} \ \ \bm{y}_i = k
\end{equation}

\begin{equation}
	\bm{\Sigma}_k = \operatorname*{cov}[\bm{x}_i] \quad \forall i \ \  \text{s.t.} \ \ \bm{y}_i = k
\end{equation}

are the maximum likelihood estimators of the class mean and covariances. The classification performance of the GMM in several classification tasks is plotted in Figure \ref{fig:gmm}, and confusion matrices are depicted in Figure \ref{fig:con}. In the figures, datum points used during training are marked with a grey cross. The predicted label is denoted by the centre of each marker and the true label is denoted by the colour of the border.  

As can be seen in the figures, the performance of the classification is strong in the simplest damage detection case (i.e. binary classification) but the performance in other scenarios is poor. It appears that the simplistic model and feature set used here is insufficiently rich to separate the classes. 

\subsection{Damage detection: Unsupervised learning}

In SHM tasks, it is common for data to lack labels pertaining to the damage condition of the structure that the data are representative of, often leading to SHM practitioners undertaking unsupervised learning approaches. In this setting, it is common to perform outlier analysis, where incoming data from the structure in an unknown condition are compared by some distance metric to data generated by a baseline "normal" condition. If outliers are observed in the incoming data, then it is believed that damage may have developed in the structure \footnote{Whether an outlier is truly representative of damage or not is strongly dependent on the quality of the feature extraction, and how sensitive to damage whilst insensitive to external influences such as operational and environmental fluctuations the chosen feature is.}. 

The principle of outlier detection is the computation of some distance metric between a test data point $\mathbf{x}_i$ and the centre of the distribution that we believe the normal condition data to be drawn from. Assuming that the data in the baseline condition follow a normal distribution, a standard distance or \textit{discordancy measure} is that of the Mahalanobis distance, defined as,

\begin{equation}
    D = (\mathbf{x}_i - \bm{\mu})^T\Sigma^{-1} (\mathbf{x}_i - \bm{\mu})
\end{equation}

where $\mu$ and $\Sigma$ are the mean and covariance of the normal condition observations. Monte Carlo sampling can then be used to determine a threshold for $D$ over which a test point $\mathbf{x}_i$ is believed to be an outlier. 

\begin{figure}
	\centering
	\includegraphics[width=\fw\linewidth]{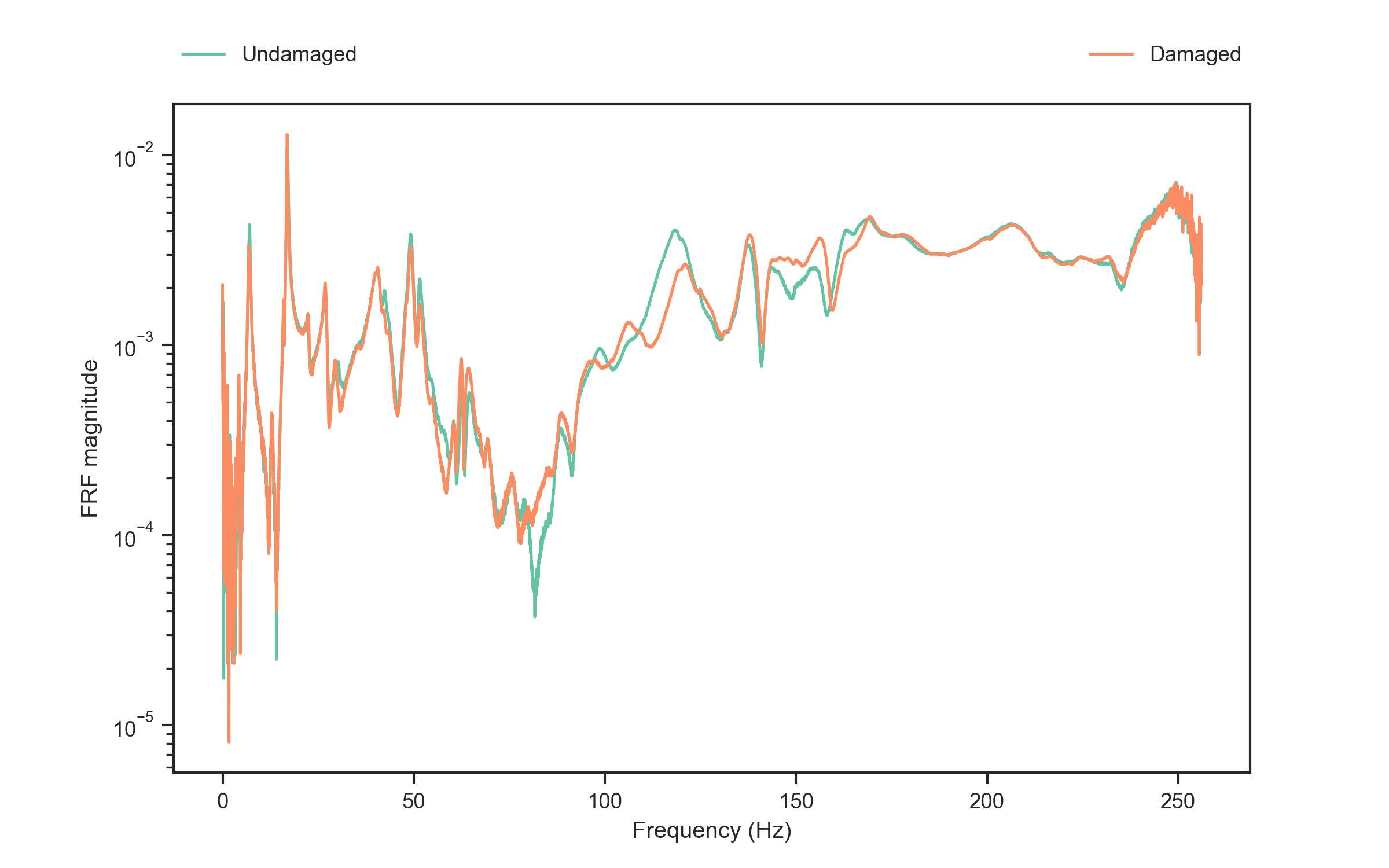}
	\caption{Comparison between undamaged and damaged (898.0g added mass at the tip) FRFs for sensor ULC-03, subject to burst-random excitation at 2.0V.}
	\label{fig:comp_1}
\end{figure}

Following the definition of a discordancy measure, all that remains to construct the outlier analysis is the selection of a feature sensitive to damage. As mentioned previously, natural frequencies are a common damage sensitive feature to use given that they will shift inline with local stiffness and mass changes that will be caused as a consequence of structural damage. To demonstrate, Figure \ref{fig:comp_1} plots the FRF extracted from the undamaged and damaged (898g added mass at the tip) condition from sensor ULC-03 (sensor positions are in reference to Figure \ref{fig:sensor_locs}) subject to burst random excitation at 2.0V. It can be seen that there is a shift in both the peak frequencies and amplitudes in the damage case relative to the undamaged, confirming a suitable damage sensitive feature. 

\begin{figure}
	\centering
	\includegraphics[width=\fw\linewidth]{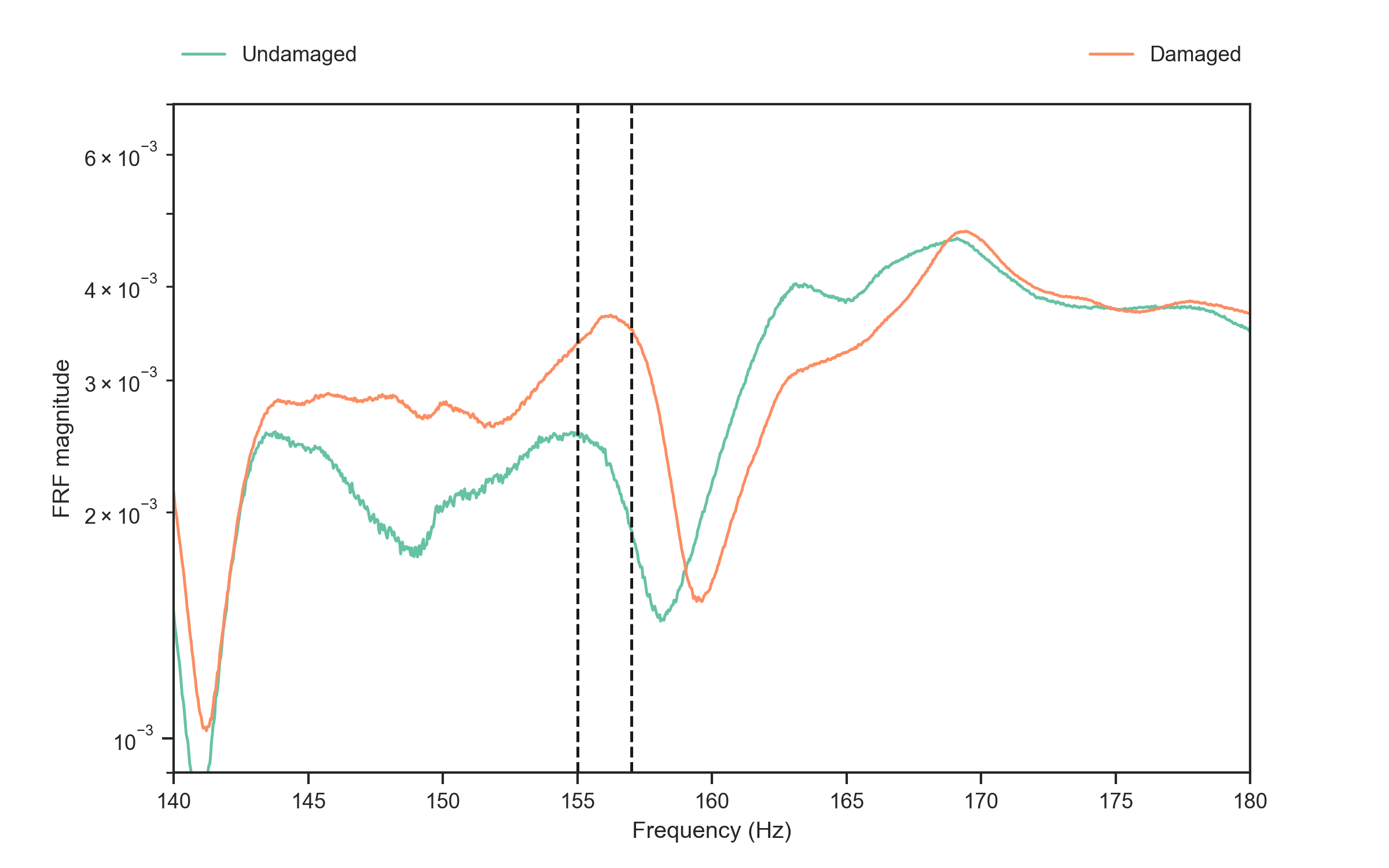}
	\caption{Zoom into damage sensitive region of the FRFs. Vertical lines indicate the frequency range used as features for the novelty detection algorithm.}
	\label{fig:comp_2}
\end{figure}

To mitigate the curse of dimensionality and avoiding taking the entire FRF as the feature vector, it is sensible to focus on just one peak of the FRF (e.g. around a natural frequency). Zooming in to the FRF around the natural frequency at 156Hz, shown in Figure \ref{fig:comp_2}, it can be seen that there is still a clear shift between the two damage cases. The feature vector is then taken to be 8 spectral lines between 155 and 157Hz. 

\begin{figure}
	\centering
	\begin{subfigure}[b]{0.49\textwidth}
		\centering
		\includegraphics[width=\textwidth]{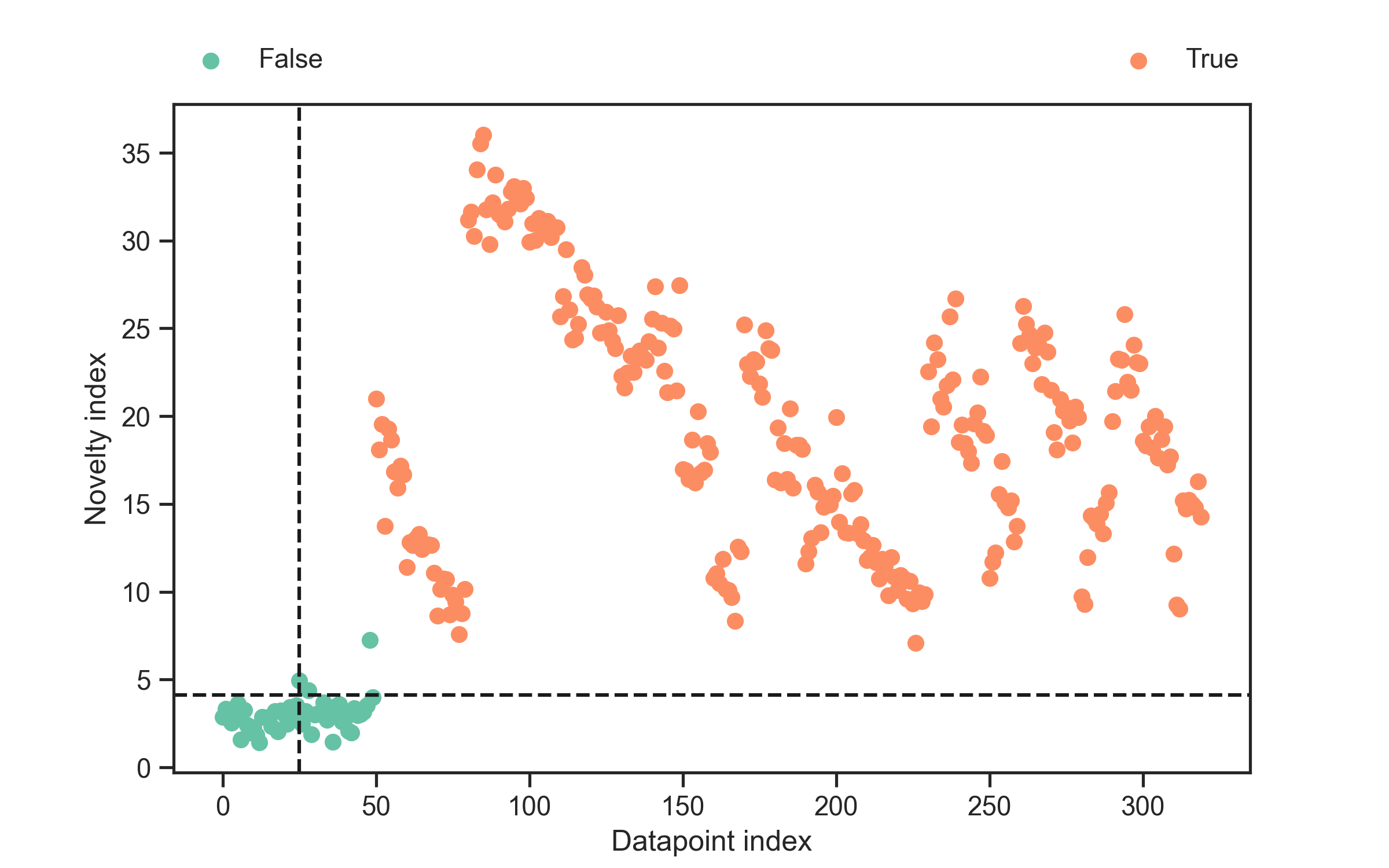}
		\caption{Presence of damage.}
		\label{fig:nov_11}
	\end{subfigure}
	\hfill
	\begin{subfigure}[b]{0.49\textwidth}
		\centering
		\includegraphics[width=\textwidth]{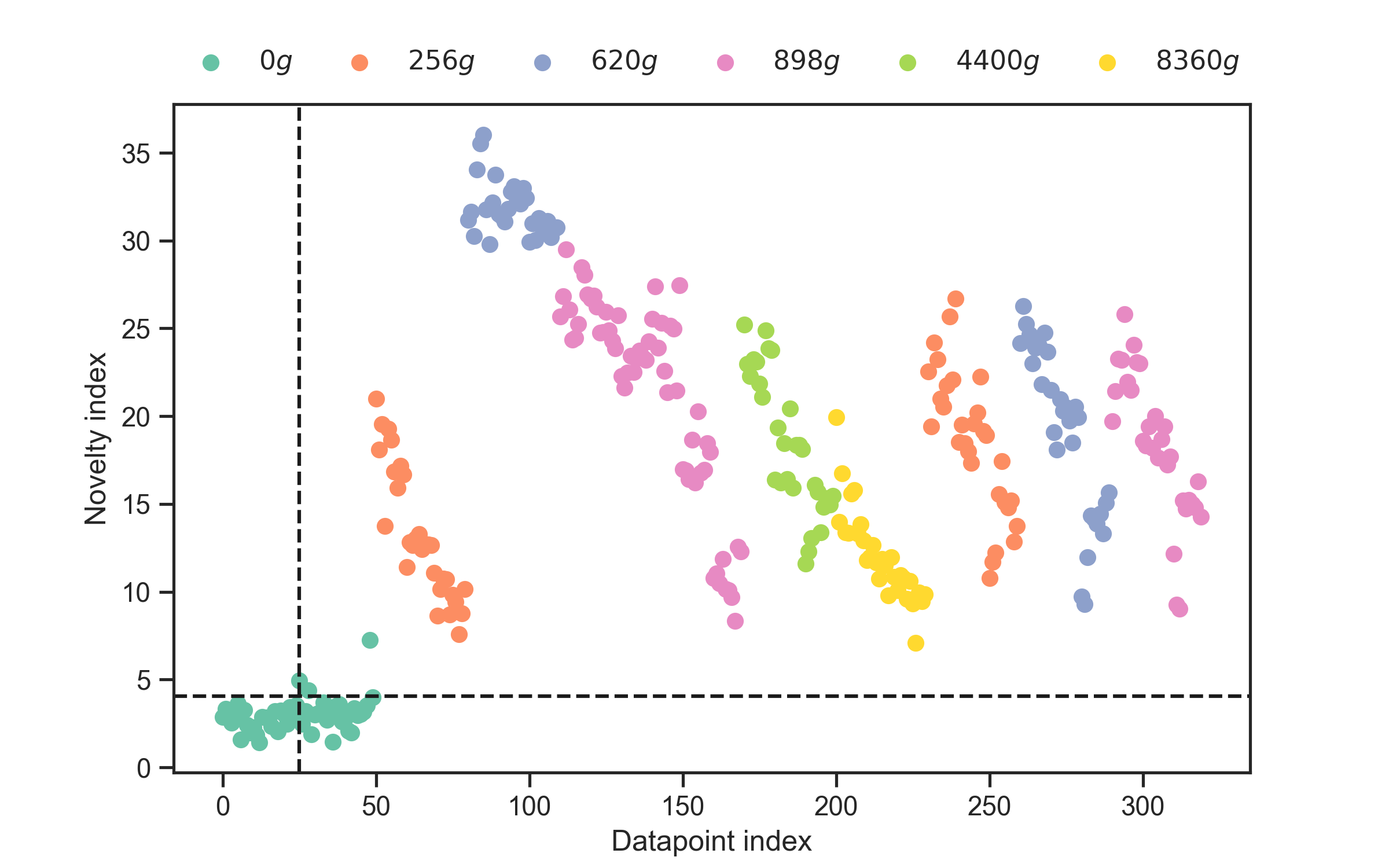}
		\caption{Added mass amount.}
		\label{fig:nov_12}
	\end{subfigure}
	\hfill
	\begin{subfigure}[b]{0.49\textwidth}
		\centering
		\includegraphics[width=\textwidth]{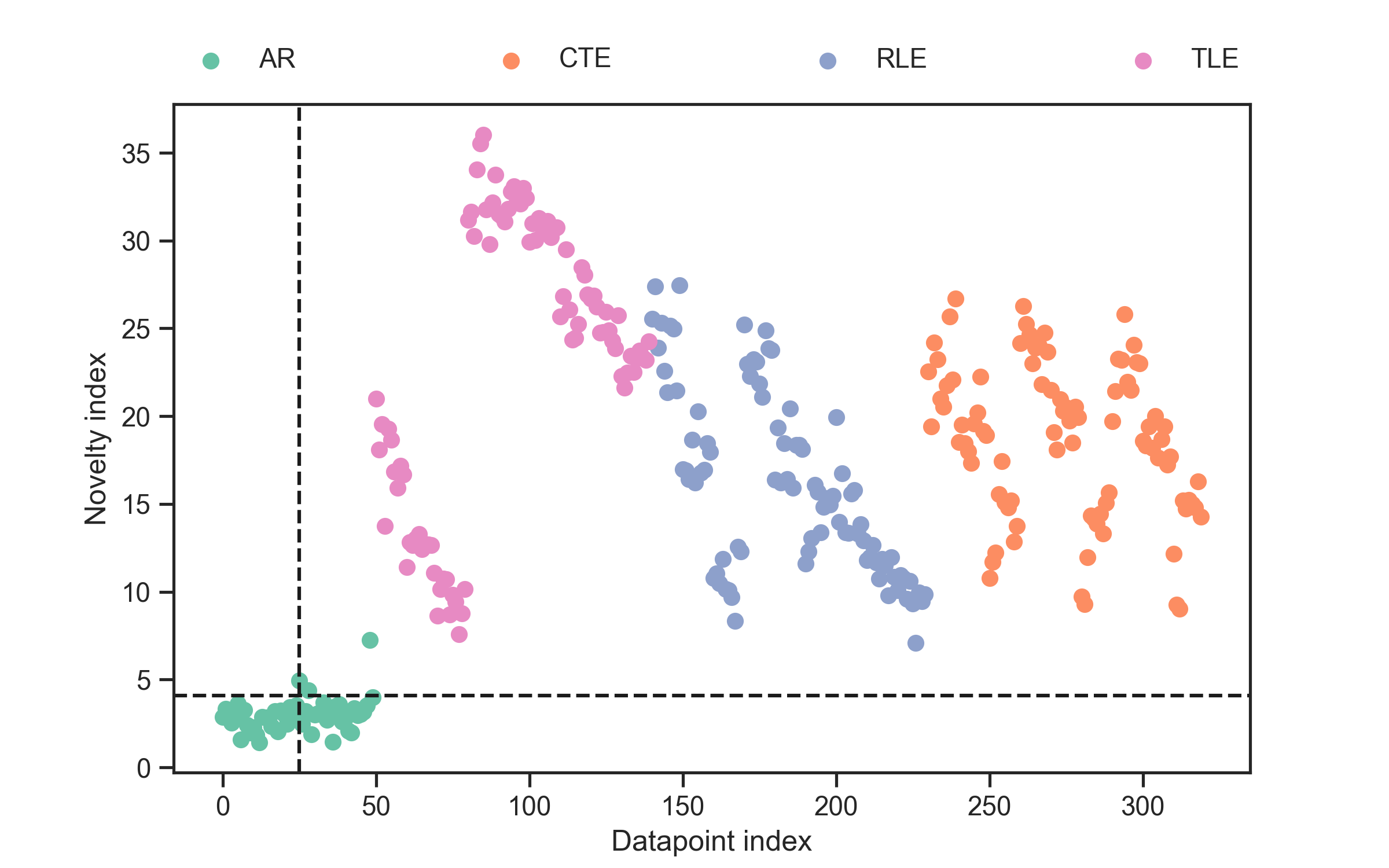}
		\caption{Damage location.}
		\label{fig:nov_21}
	\end{subfigure}
	\hfill
	\begin{subfigure}[b]{0.49\textwidth}
		\centering
		\includegraphics[width=\textwidth]{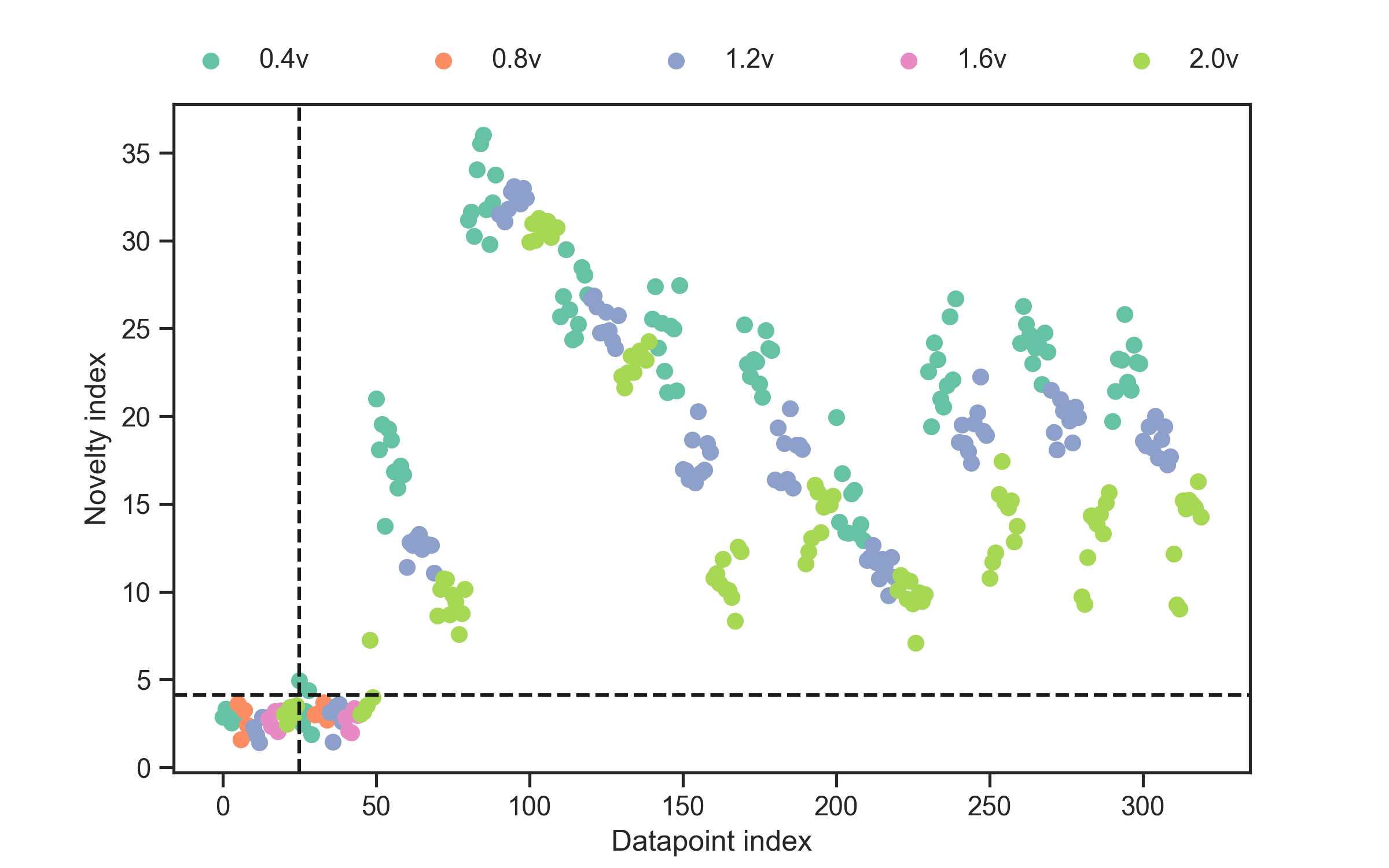}
		\caption{Excitation level.}
		\label{fig:nov_22}
	\end{subfigure}
	\caption{Results from the novelty detection algorithm, trained on half of the normal condition data. Features are spectral lines between 155 and 157Hz.}
	\label{fig:nov}
\end{figure}

Having identified a suitable input feature for the outlier analysis, it is possible to use data that is known to have been collected from the structure in of an undamaged case to form a baseline for the novelty index, which sets the threshold for whether a datum point is believed to be an outlier, and thus indicative of damage, or not. Beginning with just a comparison between data from an undamaged and damaged condition, Figure \ref{fig:nov_11} shows that the novelty detection is easily able to distinguish between the two damage classes, which includes varying levels of mass (pseudo-damage level), mass location and excitation, as shown in Figures \ref{fig:nov_12} - \ref{fig:nov_22}. With outlier analysis, however, one is limited to a binary state classification, which in this case is whether the data is representative of a healthy or damaged(anomalous) structural condition. Attempts to attempt to classify on the basis of the novelty measure online will generally be futile, as shown in Figures \ref{fig:nov_12} and \ref{fig:nov_21}, where there exists no discernible separation in the novelty measure space for data across differences in mass amount and location. 


\section{Dataset challenges}
\label{sec:challenges}

The aim of this publication has been to layout a benchmark dataset for practitioners interested in structural dynamics, SHM and digital twins. %
To this end, a set of baseline analysis has been conducted to establish what is possible with well known techniques within the field. %
As such, it seems fitting to describe here what the remaining open challenges are when considering this dataset. %
It is the hope of the authors that this (admittedly non-exhaustive) list may prompt other members of the research community to apply their state-of-the-art methods.

\subsection{Identification}

The first category of challenges in this dataset is related to the system identification of a large and complex structure. %
From the perspective of the authors, the purpose of this system identification task is to provide a model which can:
\begin{enumerate}
    \item Replicate the dynamic behaviour of the physical system through a mathematical model.
    \item Provide some level of insight into the physical mechanisms which have led to the observed response of the system.
    \item Quantify the uncertainty/confidence of the modeller, perhaps considering if the observed uncertainty is both epistemic and aleatoric.
\end{enumerate}
Any given approach may target one or more of these objectives and, ideally, all three. %
It is also worth noting that the analysis carried out should be usable in such a way that it informs further decisions, e.g.\ the results of a linear modal system identification may be used as features within an SHM scheme.

Considering the methodologies shown so far in this publication which are intended as a baseline, it is worth noting some of the major assumptions made thus far which (the authors would argue) should and can be challenged in future work. The analysis thus far has:
\begin{itemize}
    \item Considered only the \emph{low} frequency regime, this has neglected \emph{local modes}, e.g.\ the response of individual panels on the wing which may behave as thin plates.
    \item Relied on there being a \emph{modal} decomposition, such that the observed transfer function of the system is a superposition of linear modes.
    \item Extending this modal assumption, all analysis has considered the system linear, despite some observed sensitivity in the FRF to excitation amplitude.
    \item Considered the system to be completely time-invariant across a testing campaign of several weeks.
    \item The aircraft is treated directly in a reduced order form as a relatively small number of ordinary differential equations (in the decoupled modal space), compared to the overall complexity of the structure.
    \item All analysis shown in this paper has avoided a formal treatment or quantification of uncertainty, instead only point estimates are given.
\end{itemize}

Removing or replacing these assumptions and simplifications would be valuable topics for further research. %
Some potentially interesting avenues of investigation include: consideration of the best manner in which to build the low order model of the system; verification of the identification results by other algorithms; deeper investigation into the potential nonlinearity in the system and its impact.

\subsection{Prediction}

In addition to the tasks listed above which consider the identification of the system, it is also useful to imagine scenarios in which the practitioner may wish to predict certain aspects of the system behaviour. %
The extension of the identification procedures to prediction has not been shown in this work but should be an area of investigation in the future. %
The value of such predictions to the engineer would be to allow replication of (as yet) unseen behaviours; to explore the response of the system in extreme events or critical scenarios; the ability to infer the response of the system at locations which are  not measurable in operation (i.e.\ virtual sensing); to allow validation of the identified models against other collected dataset; to provide an objective means for comparison between models (especially if looking at the relative performance of physics-based and data-driven models).

It is instructive to further consider two of these tasks in greater detail. %
For the problem of virtual sensing which can be considered to reproducing data which would have been measured at some dense (by some measure) set of sensors on the basis of observations at a sparse set of sensors. %
There exist several methodologies in the literature which provide promising approaches for the virtual sensing problem. Applying these to the Hawk data presented here in future work would form a further valuable benchmark on this system. %
In addition to this there are several outstanding questions which are relevant when considering the problem of virtual sensing. %
Firstly, one could explore what would be the minimal set of sensors required to provide some bounded uncertainty/error in the locations which are being reconstructed. %
Secondly, it is important to explore which sensors are (in some sense) optimal to perform the reconstruction and how the user may choose this subset. %
Thirdly, the verification and validation of virtual sensing systems especially in terms of their reconstruction under extrapolation, e.g.\ when operating in a larger amplitude excitation regime.

Another prediction task of interest, for which this data could serve as a benchmark, is that of load estimation. %
In many settings of interest, unlike the scenario encountered in this laboratory test, it may be impractical to acquire measurements of the loads applied to the structure. %
In aerospace applications, it is very challenging to acquire the aerodynamic loads on the wings during flight. %
Similar scenarios are encountered in wind turbine or bridge applications. In all such cases, the fluid structure interaction forces are typically unavailable. %
Owing to this difficulty, in recent years, there has been some attention given to estimation of the loads alongside other properties of interest in the system. %
This is referred to as either load estimation/reconstruction or as joint input-state(-parameter) estimation \cite{Vettori2023}. %
In the context of using in service measurements for asset management, knowledge regarding the loads is highly valuable as the loads are intrinsically linked to the fatigue life of the structure. %
In practice the estimation of these loads is complicated by noise on the measurements, uncertainty regarding the dynamic system itself and a lack of full state information/observation, i.e.\ only sparse measurements are available. 

\subsection{Evaluation}

In addition to the understanding and reconstruction of the dynamic system, it is of interest to consider the follow-on tasks which this form of testing/monitoring will enable. %
One example of a task for which dynamics might be studied (and demonstrated in this paper) is that of SHM which enables automated asset management through monitoring. %
This desire for more tightly integrated modelling, incorporation of data and asset management may also be characterised, more recently, using the language of digital twins \cite{wagg2020digital,gardner2020towards}. %
In this paper tasks related to the lower levels of Rytter's hierarchy have been shown. %
For problems of damage detection, it has been seen in this work how (given insight about the form of the damage or some prior knowledge of sensitive features) it is possible to build a good quality model. %
However, doing this in the more representative context of only having access to the outputs or without good prior knowledge of the features may yet pose a challenge. %
What has been clear is that the more challenging tasks in the hierarchy of localisation, classification and extent remain very challenging based on the features extracted. %
This difficulty is present even when considering the high level of prior knowledge used in this study. %
Additionally, the interaction of the SHM procedure with, for example, the nonlinearity or other confounding influences remains to be explored. %
The topic prognosis has not been approached in this work and, although different extents of (proxy) damage are used, this would not easily be accomplished with this benchmark dataset. %
Some additional tasks which may add high value to the SHM community would be to attempt to develop a generative model of damage which would allow classifiers to be trained without examples of the previous damage cases. %
For example, this could be to reproduce the effect of damage at the trailing edge of the wing from the example data collected at the tip. %

This section has attempted to provide a non-exhaustive list of future avenues of investigation for which this dataset could prove a useful benchmark. %
It is the belief of the authors that the presented data represents a challenging full-scale example of some of the open questions within the field of structural dynamics and SHM. %
Hopefully, this discussion will stimulate further investigation of this dataset and subsequent research. 

\section{Conclusion}

In this work, a new, large-scale structural dynamics benchmark dataset is introduced. Details of the structure, experimental campaign, acquisition and control systems are presented and the data is made freely available. As well as the supplied data, this work also presents a first analysis of several dataset challenges using standard approaches in structural dynamics. For the task of system identification, a \emph{rational fraction decomposition} is applied in the frequency domain, and this is compared to an output only time domain identification in \emph{stochastic subspace identification}. 

For the task of damage identification, a \emph{supervised} machine learning approach based on \emph{Gaussian mixture models} on the principal coordinates of the natural frequencies is considered, with an \textit{outlier analysis} presented as a first benchmark as an \emph{unsupervised} method.

The analyses and results presented here are not groundbreaking and are intended to serve as a baseline, against which, more advanced methods might be compared. Some challenges which the dataset presents have been discussed, with the view to inspire readers with motivation for new approaches to structural dynamics analysis. %

\section*{Data availability}

All data and documentation as well as a python API for accessing and interacting with the data are openly available at \url{https://doi.org/10.15131/shef.data.22710040.v1}. 

\section*{Author contributions}
\textbf{MHA}: Conceptualisation (Supporting), Methodology (Lead (Experimental)), Software (Lead (Experimental)), Investigation (Lead), Data Curation (Supporting), Writing - Original Draft (Lead), Writing - Review \& Editing (Supporting)

\textbf{RM}: Conceptualisation (Supporting), Methodology (Supporting), Software (Supporting), Investigation (Supporting), Resources (Lead), Data Curation (Supporting)

\textbf{MDC}: Methodology (Lead (System ID)), Software (Lead (System ID)), Investigation (Supporting), Data Curation (Lead), Writing - Original Draft (Supporting), Writing - Review \& Editing (Supporting)

\textbf{MRJ}: Methodology (Lead (SHM)), Software (Lead (SHM)), Investigation (Supporting), Data Curation (Supporting), Writing - Original Draft (Supporting), Writing - Review \& Editing (Supporting)

\textbf{MSB}: Data Curation (Supporting)

\textbf{DW}: Resources (Supporting), Writing - Review \& Editing (Lead), Supervision (Supporting), Project administration (Lead), Funding acquisition (Lead)

\textbf{TJR}: Conceptualisation (Lead), Methodology (Supporting), Software (Supporting), Resources (Supporting), Data Curation (Supporting), Writing - Original Draft (Supporting), Writing - Review \& Editing (Supporting), Supervision (Lead), Project administration (Supporting), Funding acquisition (Supporting)

\section*{Acknowledgements}

The authors of the paper are thankful for the support of The Alan Turing Institute under the Data-Centric Engineering programme, specifically through the project Digital twins for high-value engineering applications (DTHIVE), as well as the EPSRC through grants EP/R006768/1, EP/W002140/1 and  EP/S001565/1. The experimental work made use of The Laboratory for Verification and Validation (LVV), which was funded by the EPSRC (grant numbers EP/J013714/1 and EP/N010884/1),

The authors are also extremely grateful for the stimulating and insightful discussions during an Isaac Newton Institute for Mathematical Sciences data-driven deep-dive "From physics-based to data-driven assessment of structures" in February 2023, which was supported through grant no EP/R014604/1.

Finally, the authors extend their thanks to Aidan Hughes, Brandon O'Connell, Daniel Brennan, Daniel Pitchforth, George Tsialiamanis, Jack Poole, Jacques Mclean, Joe Longbottom, Tristan Gowdridge for assistance in collecting the data. 

\bibliographystyle{elsarticle-num} 
\bibliography{main_biblio}

\end{document}